\documentclass{mn2e}
\usepackage{times}
\usepackage{amssymb}
\usepackage{epsfig}

\begin{document}
\hsize=6truein

\title[Host galaxies of luminous quasars at $z \simeq 4$]{The host galaxies and black hole-to-galaxy mass ratios of luminous quasars at ${\bf z \simeq 4}$}

\author[T.A.~Targett, et al.]
{Thomas A. Targett$^{1}$\thanks{Email: tat@roe.ac.uk}, James S. Dunlop$^{1}$, Ross J. McLure$^{1}$\\
\footnotesize\\
$^{1}$ SUPA\thanks{Scottish Universities Physics Alliance}, 
Institute for Astronomy, University of Edinburgh, 
Royal Observatory, Edinburgh, EH9 3HJ, UK}

\maketitle

\begin{abstract}
Deep $K$-band imaging of the most luminous $z \simeq 4$ quasars currently offers the earliest possible view of the mass-dominant stellar populations of the host galaxies which house the first super-massive black holes in the Universe. This is because, until the advent of the James Webb Space Telescope, it is not possible to obtain the necessary deep, sub-arcsec resolution imaging at rest-frame wavelengths $\lambda_{rest} > 4000$\AA\ at any higher redshift. We here present and analyse the deepest, high-quality $K_S$-band images ever obtained of luminous quasars at $z \simeq 4$, in an attempt to determine the basic properties of their host galaxies less than 1 Gyr after the first recorded appearance of black holes with $M_{bh} > 10^9\,{\rm M_{\odot}}$. To maximise the robustness of our results we have carefully selected two Sloan Digital Sky Survey quasars at $z \simeq 4$. With absolute magnitudes $M_i < -28$, these quasars are representative of the most luminous quasars known at this epoch but they also, crucially, lie within 40 arcsec of comparably-bright foreground stars (required for accurate point-spread-function definition), and have redshifts which ensure line-free $K_S$-band imaging. The data were obtained in excellent seeing conditions ($<0.4$-arcsec) at the European Southern Observatory on the Very Large Telescope with integration times of $\simeq 5.5$ hours per source. Via carefully-controlled separation of host-galaxy and nuclear light, we estimate the luminosities and stellar masses of the host galaxies, and set constraints on their half-light radii. The apparent $K_S$-band magnitudes of the quasar host galaxies are consistent with those of luminous radio galaxies at comparable redshifts, suggesting that these quasar hosts are also among the most massive galaxies in existence at this epoch. However, the quasar hosts are a factor $\sim 5$ smaller ($\langle r_{1/2} \rangle = 1.8\,{\rm kpc}$) than the host galaxies of luminous low-redshift quasars. We estimate the stellar masses of the $z \simeq 4$ host galaxies to lie in the range $2 - 10 \times 10^{11}\,{\rm M_{\odot}}$, and use the CIV emission line in the Sloan optical spectra to estimate the masses of their central supermassive black holes. The results imply a black-hole:host-galaxy mass ratio $M_{bh}:M_{gal} \simeq 0.01 - 0.05$. This is an order of magnitude higher than typically seen in the low-redshift Universe, and is consistent with existing evidence for a systematic growth in this mass ratio with increasing redshift (i.e. $M_{bh}:M_{gal} \propto (1+z)^{1.4-2.0}$), at least for objects selected as powerful active galactic nuclei.
\end{abstract}


\section{INTRODUCTION}

At all but the most modest redshifts, the measurement of the velocity width of broad permitted lines (e.g. H$\beta$, MgII, CIV) from unobscured Type-1 active galactic nuclei (AGN) offers the only direct way to estimate the masses of super-massive black holes (Wandel, Peterson \& Malkan 1999; Kaspi et al. 2000; McLure \& Dunlop 2001; Ho 2002; McLure \& Jarvis 2002; Willott et al. 2003; McLure \& Dunlop 2004; Vestergaard \& Peterson 2006; Peng et al. 2006a; Vestergaard et al. 2008; Willott et al. 2010). For this reason, determining the properties of AGN host galaxies remains of crucial importance for exploring the relationship between black-hole and galaxy evolution over cosmic history, and in particular for testing alternative theories for the origin of the now well-established proportionality between black-hole and galaxy bulge mass found in the local/present-day Universe (Kormendy \& Richstone 1995; Magorrian et al. 1998; Silk \& Rees 1998; Gebhardt et al. 2000; Merritt \& Ferrarese 2001; McLure \& Dunlop 2002; Tremaine et al. 2002; Bettoni et al. 2003; Marconi \& Hunt 2003, Merloni et al. 2010; Jahnke \& Maccio 2011).

Quasars, as the most extreme of AGN, offer potentially the greatest challenge to theory, especially at high redshift where, for example, black holes with masses $M_{bh} > 1 \times 10^{9}\,{\rm M_{\odot}}$ already appear to have been in place less than 1 billion years after the Big Bang (Fan et al. 2001, 2003; Willott et al. 2007; Mortlock et al. 2011). However, extracting reliable information on the host galaxies of these overwhelmingly bright AGN presents a formidable technical challenge. 

At low redshift ($z < 0.4$) this challenge was effectively met when the first refurbishment of the {\it Hubble Space Telescope} ({\it HST}) enabled optical imaging of the necessary depth {\it and} exquisite+stable angular resolution to allow the scalelengths and morphologies of low-redshift quasar hosts to be reliably determined for the first time (e.g. Disney et al. 1995; McLure et al. 1999; Dunlop et al. 2003; Floyd et al. 2004). Indeed, the results of these deep {\it HST} optical imaging programs of quasar hosts, when combined with ``virial'' measurements of the black-hole masses in the same objects, were instrumental in demonstrating that these most active objects displayed the same black-hole:host-galaxy mass proportionality as found in quiescent galaxies (McLure \& Dunlop 2002).

However, extending the effective study of quasar host galaxies out to higher redshifts, and in particular into the ``quasar epoch'' at $z > 2$, has proved to be extremely difficult. There are a number of reasons for this. First, there is the obvious cosmological dimming of the host-galaxy light. Second there is the necessity of observing in the near-infrared, in order to continue to sample the rest-frame optical light of the host galaxy (both for comparison with low-redshift studies, and to avoid the host-galaxy light being completely swamped by the UV-bright quasar). Some success in detecting the hosts of quasars out to $z \simeq 2$ was achieved with the NICMOS infrared camera on {\it HST} (Kukula et al. 2001, Ridgway et al. 2001; Peng et al. 2006b) but, even with this high-resolution near-infrared imaging, obtaining reliable measurements of host-galaxy luminosities and scalelengths proved extremely difficult. We now know that this was almost certainly in part due to the fact that, in general, massive galaxies at high-redshift have since been discovered to be generally much more compact than their low-redshift counterparts (Daddi et al. 2005; Trujillo et al. 2006, 2007; Longhetti et al. 2007; Zirm et al. 2007; Cimatti et al. 2008; van Dokkum et al. 2008; Buitrago et al. 2008; Szomoru et al. 2010; Targett et al. 2011). If this is also true for quasar hosts, then the separation of galaxy and nuclear light in quasars at $z \ge 2$ will inevitably be even more problematic than was perhaps first anticipated.

Nonetheless, these difficulties have not deterred several groups from attempting to repeat this experiment at redshifts as high as $z \simeq 4 - 5$. Such studies have been encouraged (at least in part) by the advent of active/adaptive optics on ground-based 8-m class telescopes, and by the fact that $z \simeq 4$ is the natural limit for ground-based studies of quasar host galaxies if one hopes to detect host-galaxy light longward of the 4000\AA/Balmer break in the $K$-band. 

However, perhaps unsurprisingly, existing studies of the host galaxies of quasars at $z \simeq 4$ have met with very limited success. For example, using {\it HST} NICMOS $H$-band data, Peng et al. (2006b) attempted to measure the luminosities and sizes of the hosts of two gravitationally-lensed quasars at $z=4.1$ and $z=4.5$ via two-dimensional modelling. Although reliable host-galaxy luminosities were claimed, the {\it HST} images sampled light shortward of $\lambda_{rest} = 4000$\,\AA\ and, in addition, the complexity of the lensing model led the authors to conclude that the extracted morphological parameters could not be trusted. In an alternative ground-based approach, Hutchings (2003, 2005) observed seven quasars at $z \simeq 5$ with the Gemini 8-m telescope in the $J$, $H$, and $K$-bands. Using point-spread function (PSF) subtraction, host-galaxy luminosities were estimated from the PSF-subtracted images, although no evidence of smooth centrally-concentrated host galaxies was found and, again, this time due to the decision to observe quasars at $z \simeq 5$, this imaging sampled rest-frame wavelengths $\lambda_{rest} < 4000$\,\AA, limiting the usefulness of the extracted luminosities. Most recently, McLeod \& Bechtold (2009) observed 34 $z \simeq 4$ quasars in the $K$-band with the Magellan I and Gemini North telescopes. However, despite the use of well-controlled PSF subtraction, only four host galaxies were detected in their deepest and sharpest data, at an average quasar redshift $\langle z \rangle = 3.8$. Host-galaxy luminosity and size estimates were attempted from the PSF-subtracted images of these 4 quasars, although the authors noted that the surface-brightness limit of these data was really of inadequate depth for the accurate recovery of host-galaxy sizes. The deepest imaging obtained by these authors ($\simeq 3.5$ hours per source on 8-m class telescopes) would thus seem to represent a minimum requirement for the discovery and study of quasar host galaxies at $z \simeq 4$.

In this paper we present and analyse even deeper ($\simeq 5.5$ hours per source) European Southern Observatory (ESO) 8.2-m Very Large Telescope (VLT) $K_S$-band imaging of two $z \simeq 4$ Sloan Digital Sky Survey (SDSS) quasars. These two quasars were carefully chosen to have a redshift which ensures line-free imaging of the host-galaxy stellar population at $\lambda_{rest} > 4000$\,\AA, and to have a star of comparable brightness within an angular radius of 40 arcsec, in order to provide a robust, high signal:noise ratio, on-detector representation of the VLT $K_S$-band PSF over the complete duration of the host-galaxy imaging. This care in quasar target selection, combined with deliberate (flexibly scheduled) use of only the very best seeing conditions (and the deliberate avoidance of adaptive optics), has enabled us to achieve robust detections of the host galaxies of both quasars, despite the fact that, with $M_i < -28$, these objects are representative of the most luminous/massive black holes in existence at this (or indeed any) epoch. 

This paper is structured as follows. In Section 2 we describe our observing strategy, our target selection, and summarise the new near-infrared data obtained with the VLT. Next, in Section 3 we explain the PSF subtraction technique used to separate host and nuclear light. Then, in Section 4 we describe the determination of host-galaxy properties and the estimation of the central black-hole masses. Finally, in Section 5 we attempt to place our results in the context of studies of other galaxy populations at both high and low redshift, and briefly explore the consequences of this study for the inferred cosmic evolution of the black hole-to-galaxy mass ratio. Throughout we adopt a cosmology with $H_{0}=70$\,km\,s$^{-1}$\,Mpc$^{-1}$, $\Omega_{m}=0.3$, and $\Omega_{\Lambda}=0.7$. Unless otherwise noted, we report all magnitudes in the Vega system to ease comparison with previous studies, but our $K_S$-band photometry can be simply converted to the AB magnitude system via $K_{S,AB} = K_{S,Vega} + 1.85$.

\section{Observations and data reduction}

\subsection{Instrument}

The deep $K_S$-band imaging presented here was performed with the 8.2-m VLT (located on Cerro Paranal in the Atacama Desert, northern Chile), using the Infrared Spectrometer And Array Camera (ISAAC), a 1024 $\times$ 1024 Hawaii Rockwell array with a pixel scale of 0.15 arcsec. This pixel scale provided the necessary good sampling of even the very best seeing disc, while still providing sufficient field-of-view to include the required nearby field stars for accurate on-image PSF characterisation.

\subsection{Observational strategy}

The reliable detection of host-galaxy light around luminous quasars is difficult at all redshifts, but especially at high redshift, due to the small angular size of any reasonable anticipated host galaxy, the faintness of any potential host galaxy relative to the scattered nuclear light in the wings of the PSF, and the fact that the rest-frame optical light has been redshifted to the near-infrared.

Predicting surface brightnesses (and hence exposure times) for high-$z$ quasar host galaxies is complicated by unknown amounts of stellar and morphological evolution to offset against the cosmological dimming and anticipated $k$-corrections. In an attempt to best define our observational strategy we generated synthetic VLT/ISAAC exposures of a $z=4$ quasar with a nuclear/host ratio of 15, 0.6-arcsec seeing, and background-limited noise. The synthetic quasar had a total $K_S$-band apparent magnitude of $K_S=16.5$ (consistent with the anticipated $K_S$-band magnitudes of our target SDSS quasars), comprising a nuclear point source with $K_S=16.8$ and a de Vaucouleurs host galaxy with $K_S=19.7$ (consistent with existing studies of the $K-z$ relation for massive galaxies). For the simulated host galaxy we assumed a half-light radius of 5\,kpc. Separate PSFs were used to construct the synthetic quasar and to carry out nuclear/host separation, with both PSFs obtained from the publicly-available ISAAC image of the GOODS-South field, on which they were $\sim30$-arcsec apart (thus accurately reflecting our observing strategy of using a nearby star to characterise the on-image PSF). In order to separate host and nuclear components and reliably determine total luminosities for such galaxies, we found that we needed to follow the surface brightness profile out to $\sim 3$ galaxy half-light radii. This required integrating down to a $3\sigma$ surface brightness level of $\mu_K \simeq 25$\,mag\,arcsec$^{-2}$. The VLT ISAAC Exposure Time Calculator indicated that we could achieve this level in $\simeq 5$ hours on source, and so this was set as the minimum on-source exposure time for our observations.

\subsection{Target selection}

For the reasons explained above, we wished to obtain emission-line free images of the most luminous quasars at $z \simeq 4$. Contamination of the observed $K$-band light by [OIII] and H$\beta$ line emission is avoided by insisting on $z > 3.95$, and so we aimed to select quasar targets at only slightly higher redshift to ensure that the $K_S$-band filter was still essentially filled by the continuum light starting at $\lambda_{rest} \simeq 4000$\AA. We confined our attention to the most luminous quasars in the SDSS quasar catalogue at these redshifts, corresponding to absolute magnitudes $M_i < -28$. These objects are rare; the final SDSS quasar catalogue contains only $\simeq 200$ such quasars in the redshift range $4 < z < 5$ over the full $\simeq 9380$\,deg$^2$ of the SDSS quasar survey (Schneider et al. 2010). They thus have a comoving number density of $\simeq 2$ per comoving Gpc$^3$ at $z \simeq 4$, which corresponds to $\simeq 1$ object in the cosmological volume simulated in the ``Massive Black'' simulation recently discussed by Di Matteo et al. (2011) (see Section 5). 

Finally, we further limited our observational options to quasars from within this sample which happened to have at least one bright star within a radius of 40 arcsec, in order to provide a high signal:noise ratio representation of the real form of the point-spread-function (PSF) on the final science image. It was deemed essential that the nearby stars were at least as bright as the quasar at $K_S$-band, but not so bright as to risk saturation or non-linearity in the background limited imaging. This final rather stringent requirement reduced our potential target list to less than 10 quasars.

\subsection{Data}

Our insistence on long ($> 5$\,hr) integrations in the very best seeing conditions meant that we only obtained the required deep imaging for two of the potential target quasars, but all data were taken in photometric conditions and consistent high-quality ($<0.4$-arcsec) atmospheric seeing. Our observations of these two $z \simeq 4$ quasars are summarised in Table 1.

\begin{table}
 \begin{center}

\caption{Basic details of the VLT ISAAC $K_S$-band observations of the two $z \simeq 4$ SDSS luminous quasar targets. Column 1 lists the target names. Column 2 gives the total exposure times in hours. Column 3 provides the measured 4-arcsec diameter aperture $K_S$-band magnitude (Vega) for each quasar. Finally, Column 4 lists the spectroscopic redshifts of the quasars as given in the final SDSS quasar catalogue (Schneider et al. 2010).}

\begin{tabular}{lccc}

\hline
Source & Exp. time / hr & $K_S$ (Vega) & z\\
\hline
SDSS J131052.51$-$005533.2 & 5.9 & 16.50$\pm$0.04 & 4.1567\\
SDSS J144617.35$-$010131.1 & 5.1 & 16.70$\pm$0.05 & 4.1608\\
\hline
\end{tabular}

 \end{center}
\end{table}

\subsection{Data reduction}

Data reduction was performed using standard {\small{IRAF}} packages. Dark frames of equal integration time to the science data were taken on all nights. Each dark-subtracted image was divided by a normalised flat-field, derived though the scaled sigma-clipped median combination of neighbouring science frames observed within 20 minutes of each image. Registering of the offset frames was performed using the brightest stars in the images. A map of the bad pixels in the array was obtained and used to exclude these pixels during image combination. Median filtering was used to reject cosmic rays and produce an initial reduced image which was then processed to create a mask of all source flux detected in the mosaiced data. The science frames were then reprocessed to create an improved flat-field, and then the data were re-reduced to produce the final science sub-images. As accurate matching of both the quasar and PSF image quality is of critical importance in this work, separate mosaics were created for the quasars and for the PSF stars using offset shifts derived from each object individually. Final image registration was performed using a Spline3 interpolation in the IRAF routine IMSHIFT.

\section{PSF SUBTRACTION}

Because of the expected overwhelming dominance of the nuclear light, separation of nuclear and host-galaxy light was first performed using scaled PSF subtraction. We subtracted an appropriately-scaled version of the on-image stellar PSF from the quasar image to produce the minimum residual which still displays a monotonically declining surface-brightness profile (see Fig. 1), thus providing a robust estimate of the minimum level of host-galaxy flux. This method clearly represents an over-subtraction of the nuclear PSF, but at least ensures exclusion of all nuclear light from the subtracted residual. The likely extent of any over-subtraction of host-galaxy light is quantified via the host-galaxy modelling presented in Section 4.

\begin{figure*}
 \begin{tabular}{cc}

 \epsfig{file=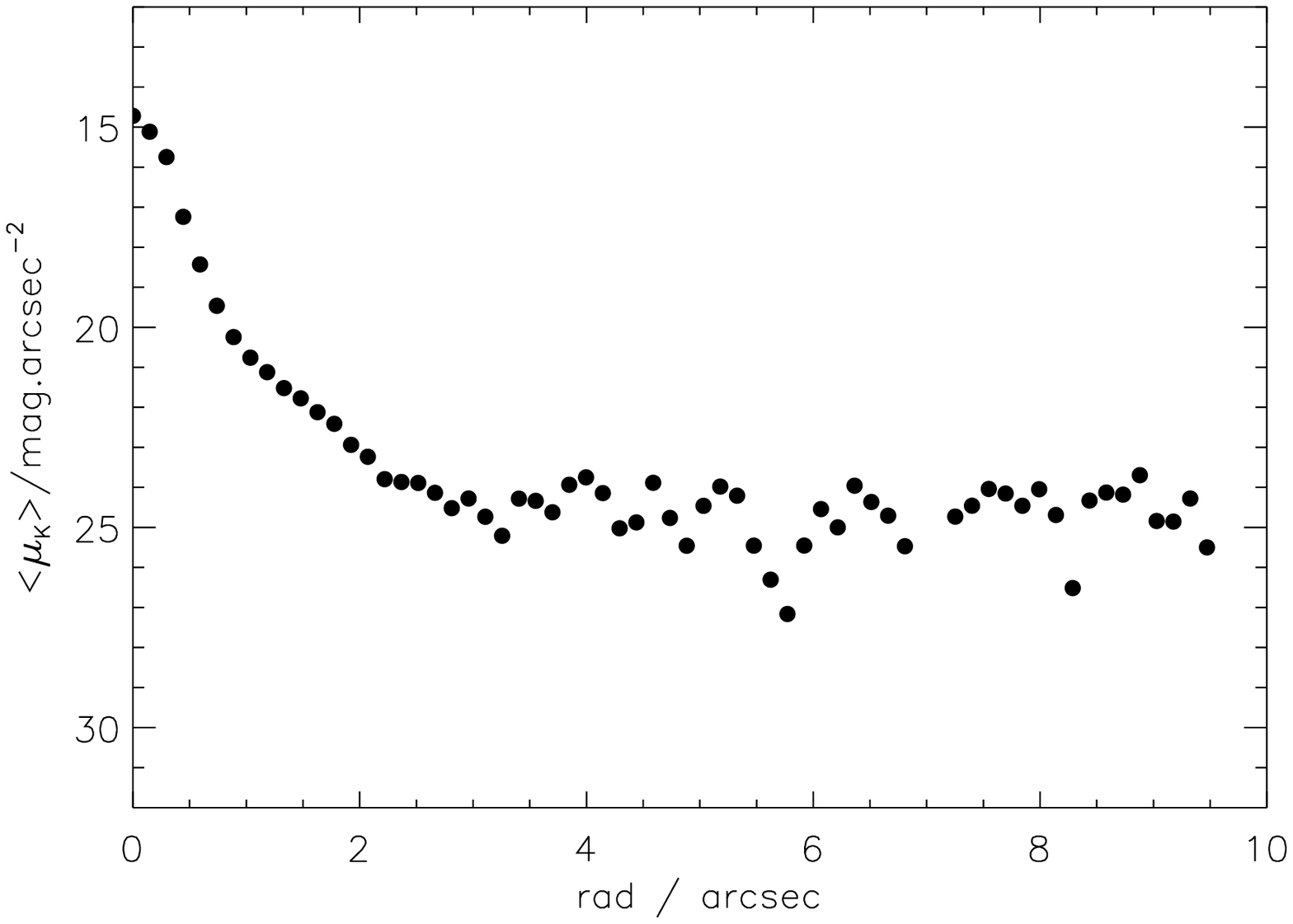,width=0.40\textwidth}&
 \epsfig{file=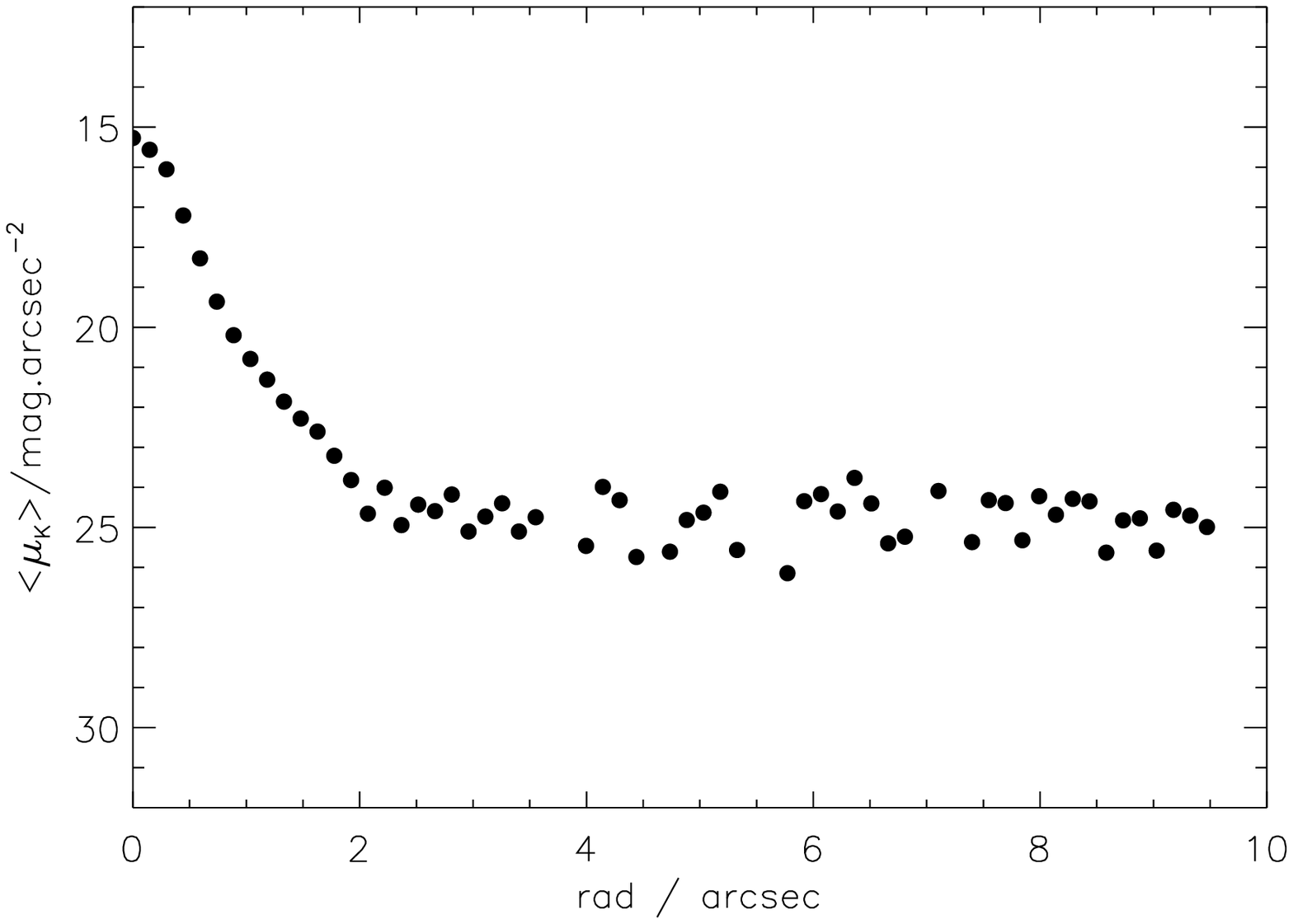,width=0.40\textwidth}\\

 \epsfig{file=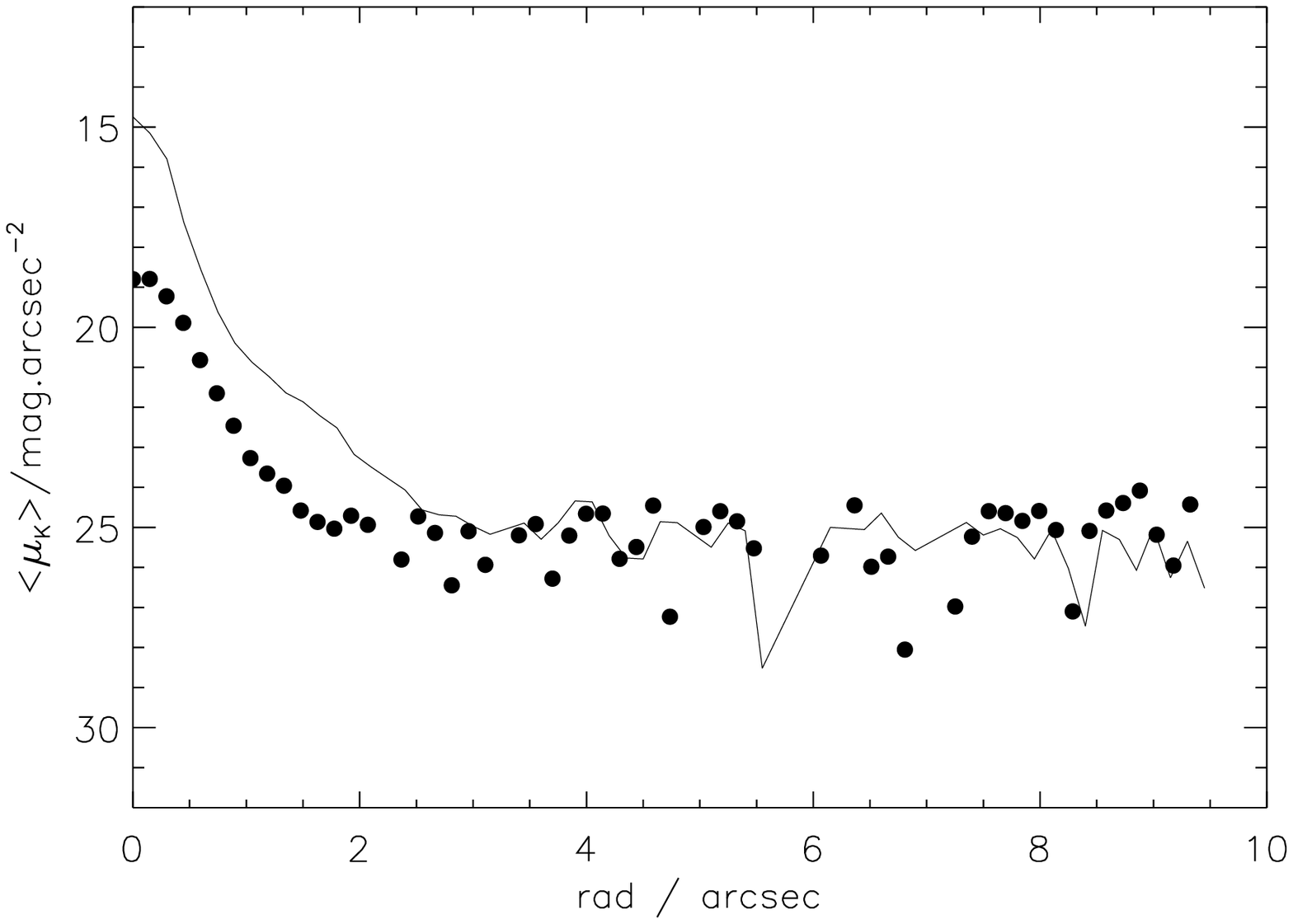,width=0.40\textwidth}&
 \epsfig{file=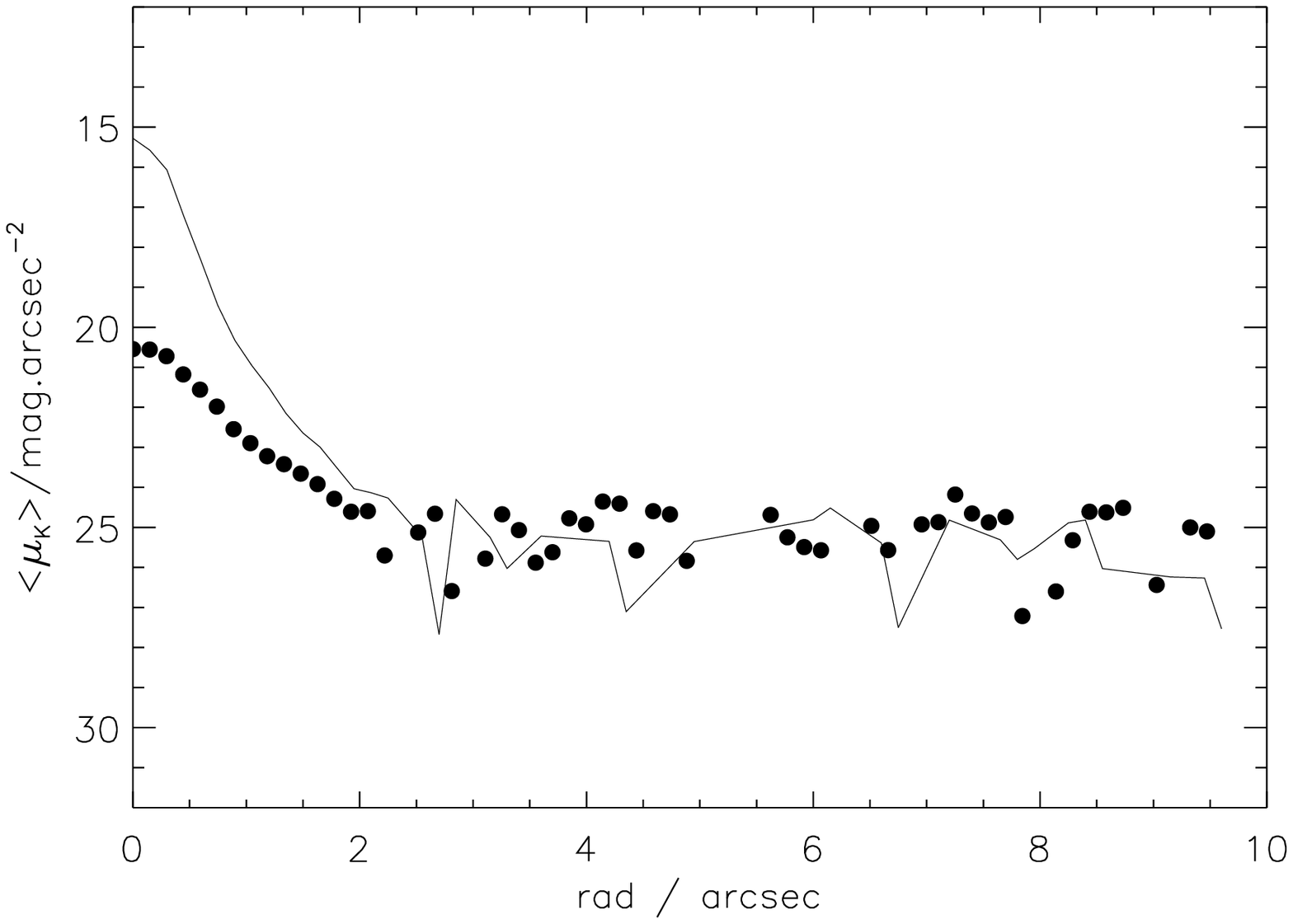,width=0.40\textwidth}\\

 \end{tabular}

 \caption{Raw quasar (top) and normalised PSF (solid line) plus PSF-subtracted monotonically-declining host-galaxy (bottom) $K_S$-band surface brightness profiles for SDSS J131052.51$-$005533.2 (left) and SDSS J144617.35$-$010131.1 (right).}

\end{figure*}

The $K_S$-band images of the two quasars, PSFs, and monotonic PSF-subtracted residuals are shown in Fig. 2. Several factors complicate determination of the exact level of PSF to subtract, where small differences between the shape of quasar and stellar PSFs can create an artificial residual. In order to check that the quasar-PSF subtracted residuals represent host-galaxy flux and not such anomalies, PSF-PSF subtractions were performed using alternate stars on the science images, with the scaling of the subtracted PSF again adjusted to the point where the residual (just) displayed a monotonically decreasing radial surface brightness profile. Reassuringly, the resulting residuals from these tests (an example of which is shown in Fig. 2) were always much smaller than (and qualitatively very different from) the putative quasar host galaxies. The results of these tests were also used to determine the typical photometric error introduced by the small but inevitable mismatches between the PSFs at different points in the ISAAC image. Finally, there is also inevitably some uncertainty when determining the precise level of PSF normalisation required for subtraction to yield the minimum monotonically declining residual, with a narrow range of scaled PSFs producing acceptable monotonic residuals. 

\begin{figure*}
 \begin{tabular}{ccc}

  \epsfig{file=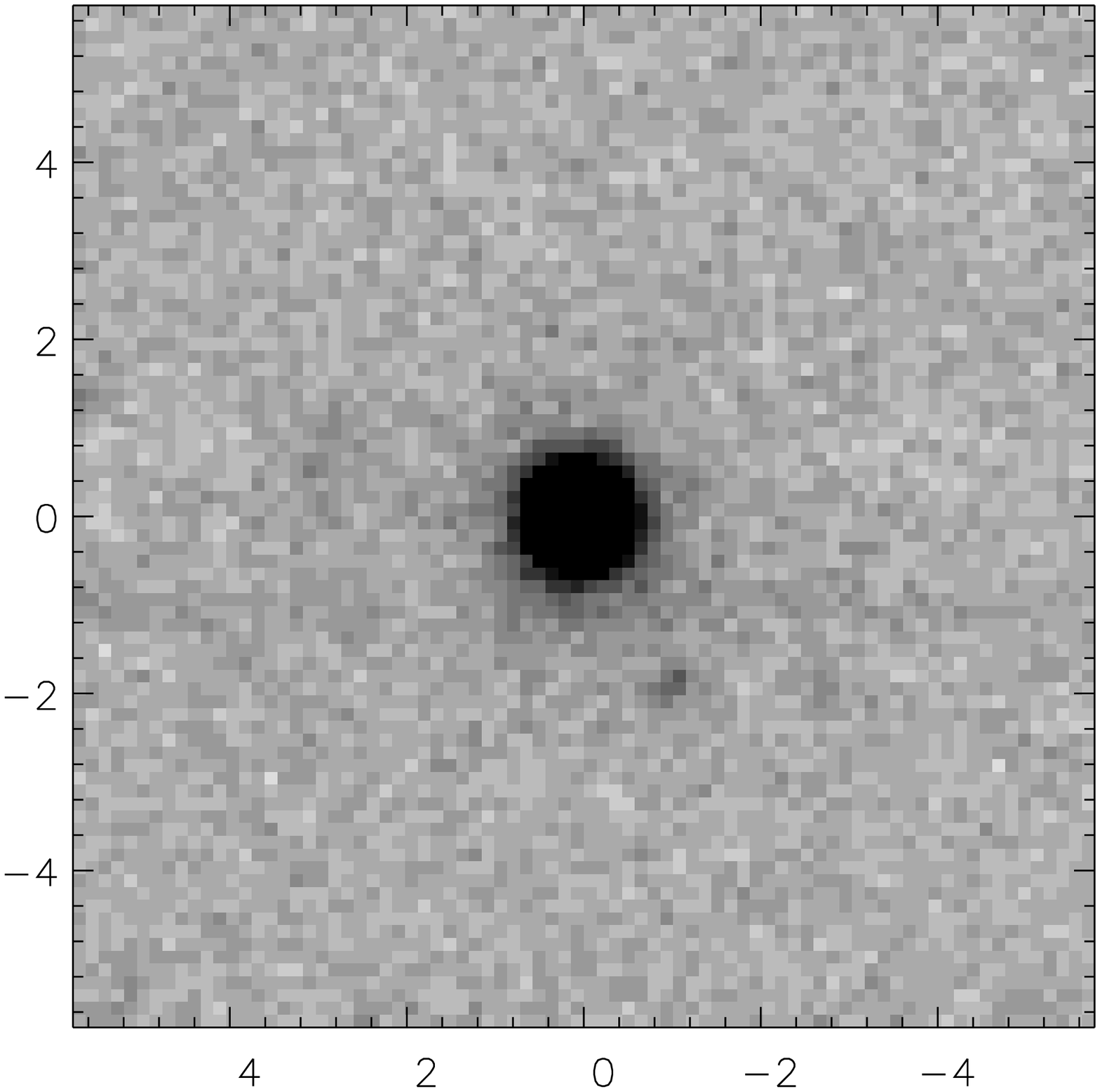,width=0.30\textwidth}&
  \epsfig{file=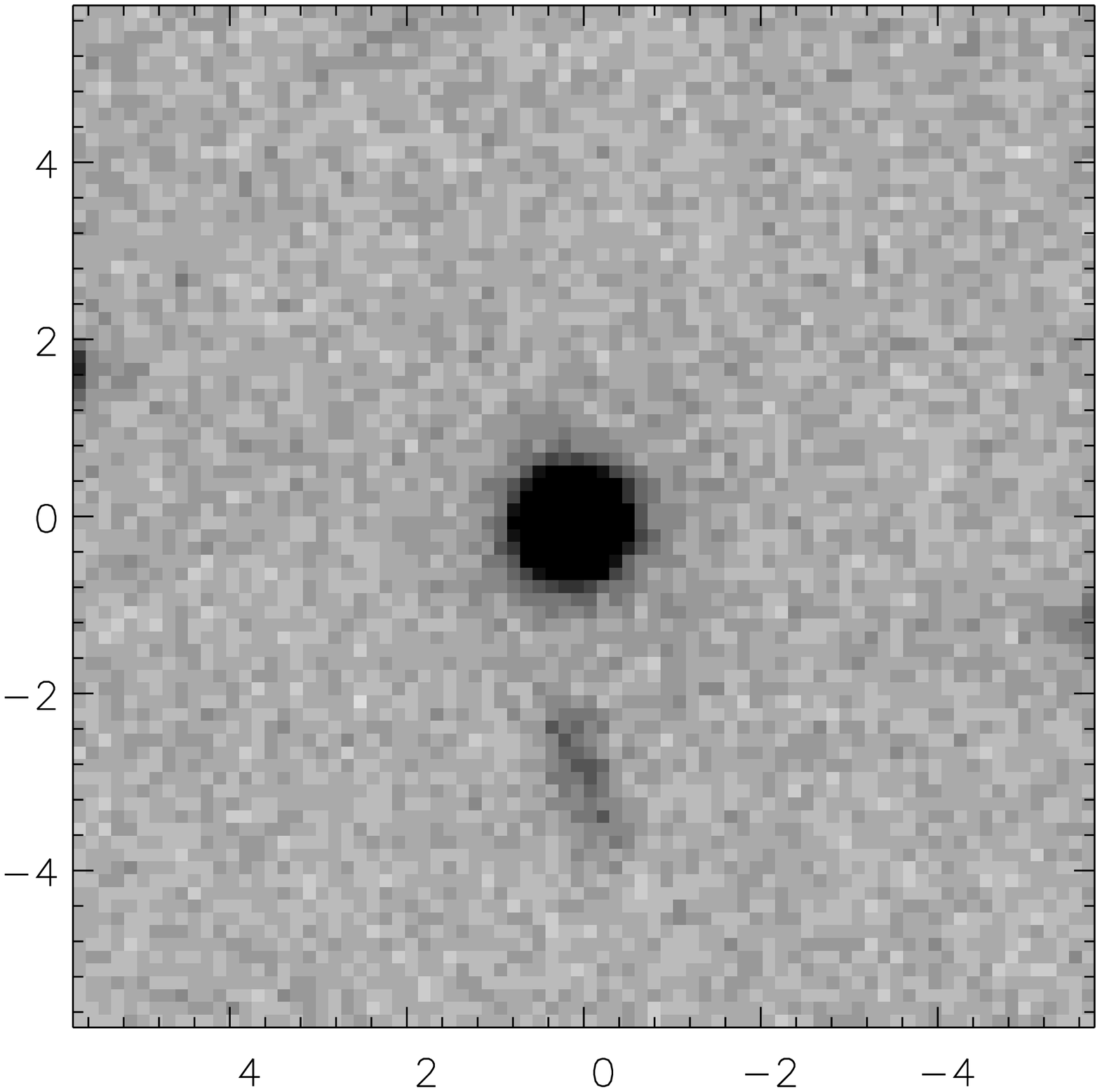,width=0.30\textwidth}&
  \epsfig{file=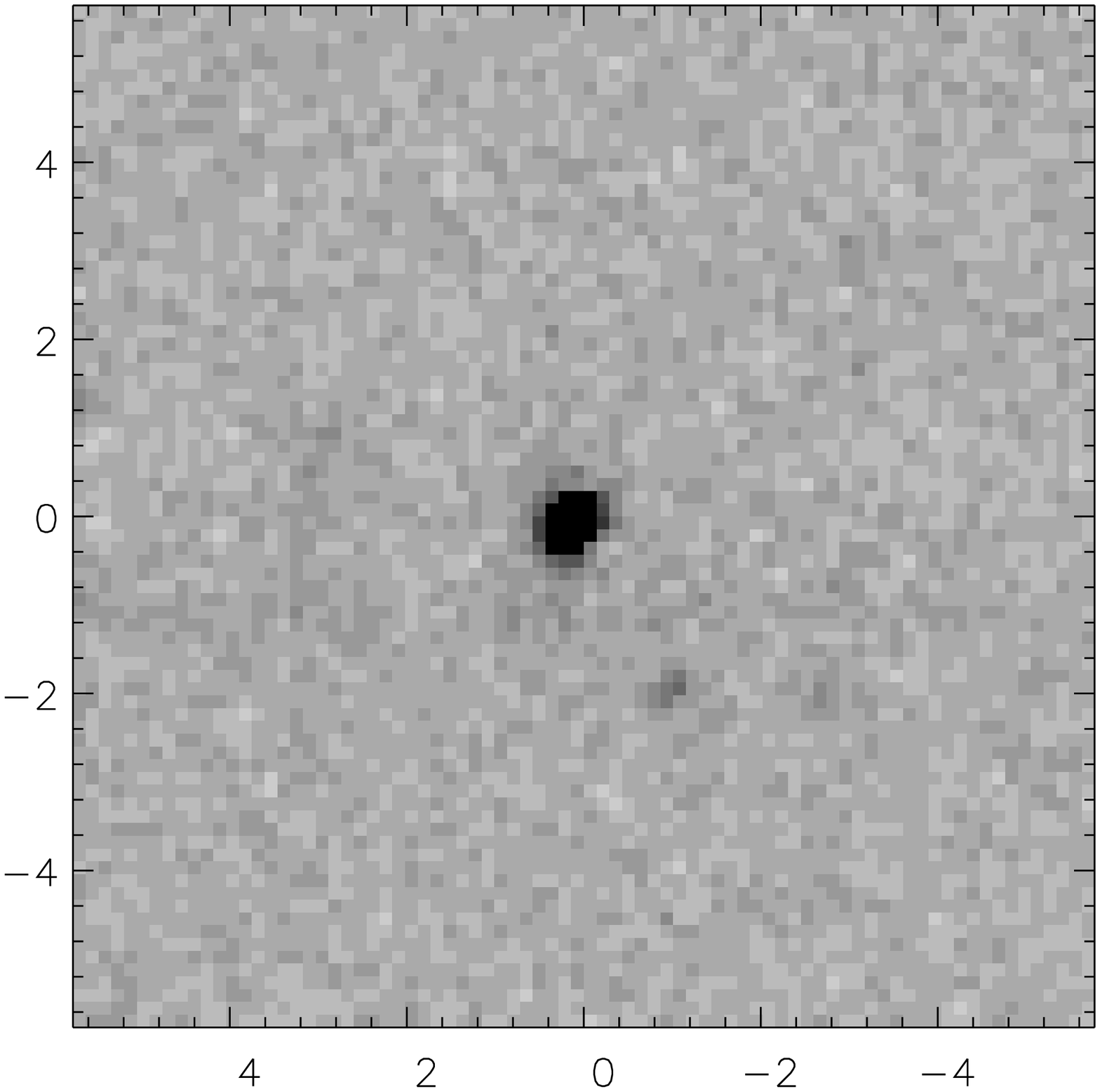,width=0.30\textwidth}\\
  \epsfig{file=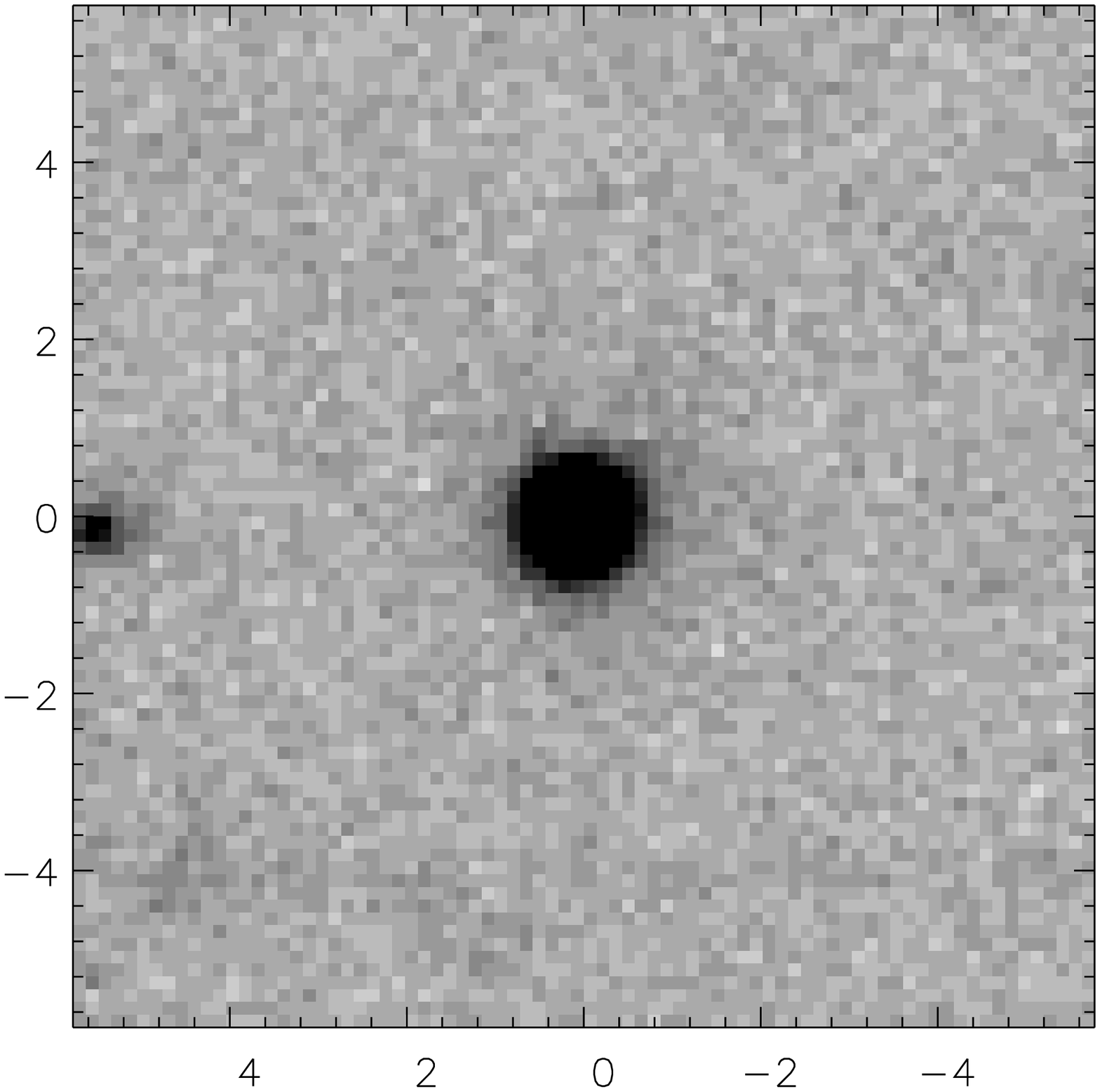,width=0.30\textwidth}&
  \epsfig{file=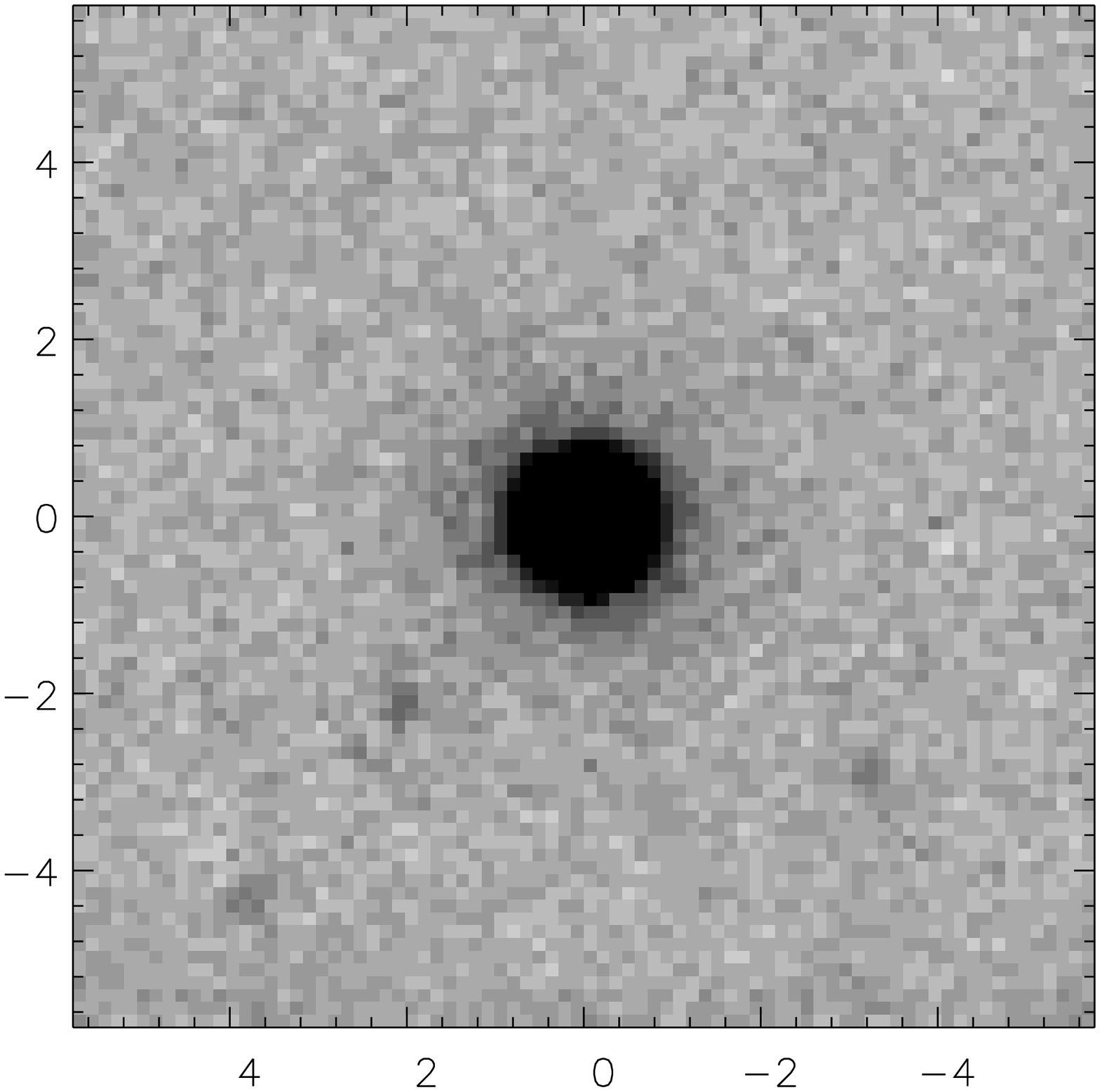,width=0.30\textwidth}&
  \epsfig{file=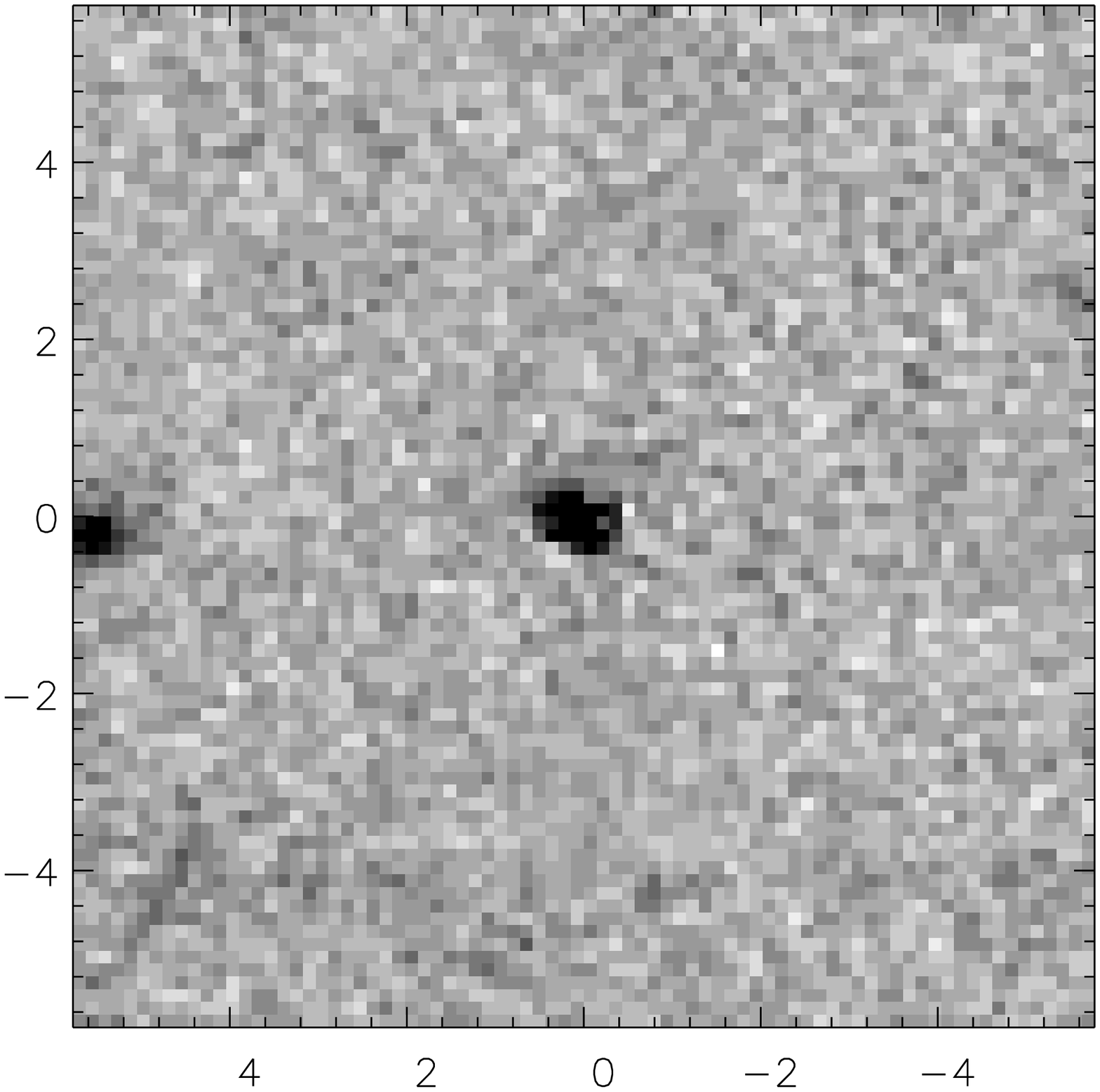,width=0.30\textwidth}\\
  \epsfig{file=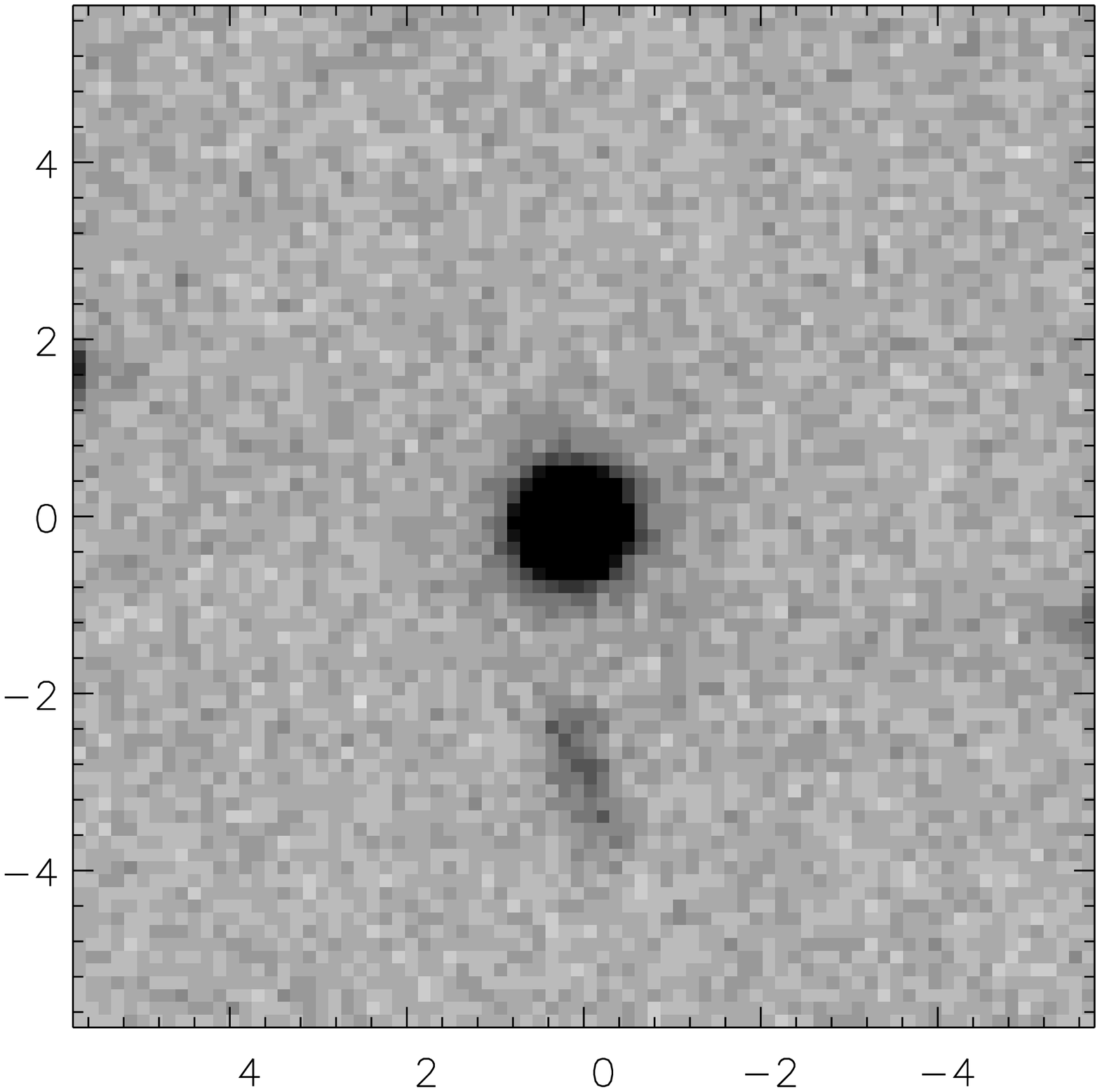,width=0.30\textwidth}&
  \epsfig{file=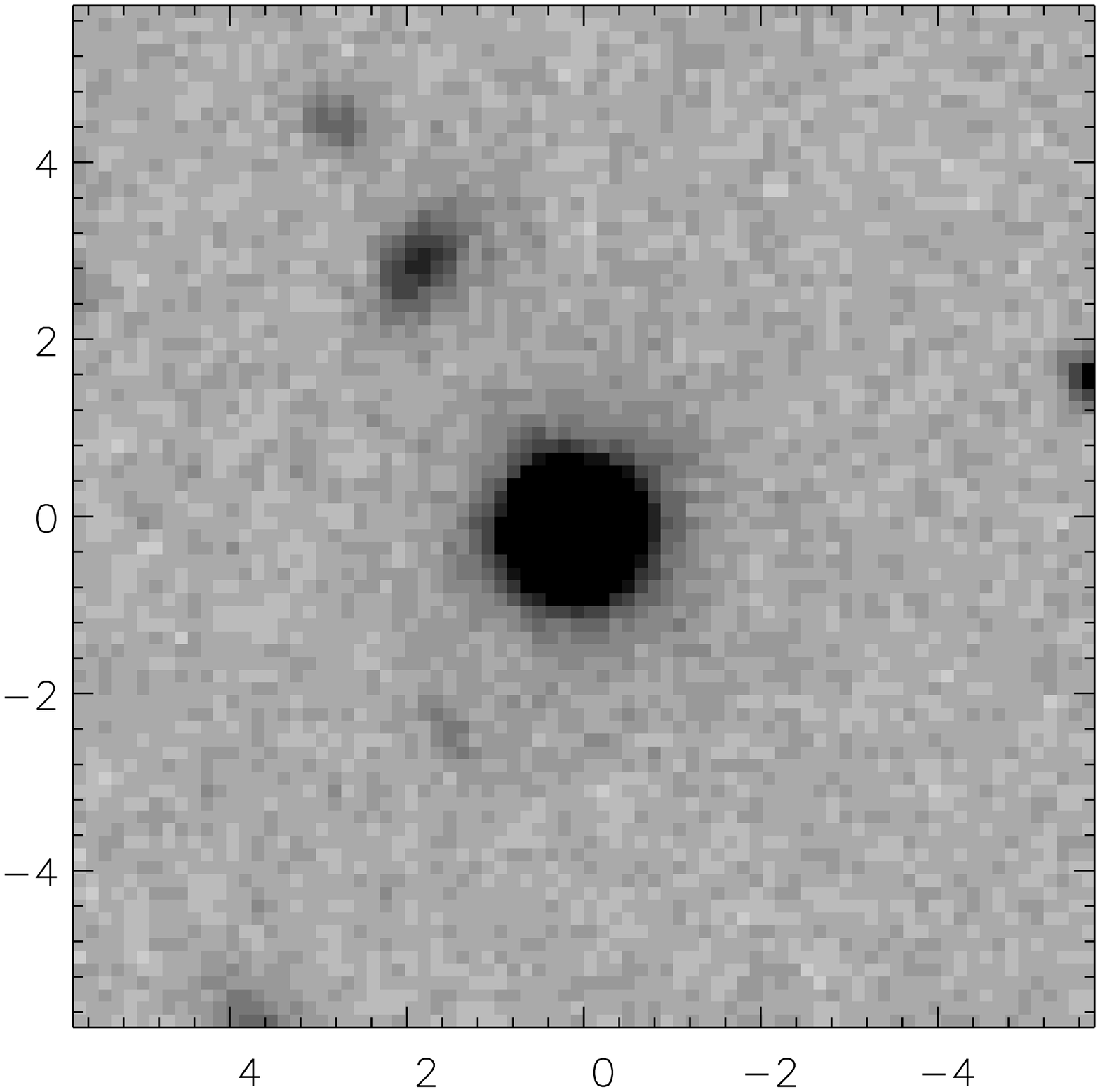,width=0.30\textwidth}&
  \epsfig{file=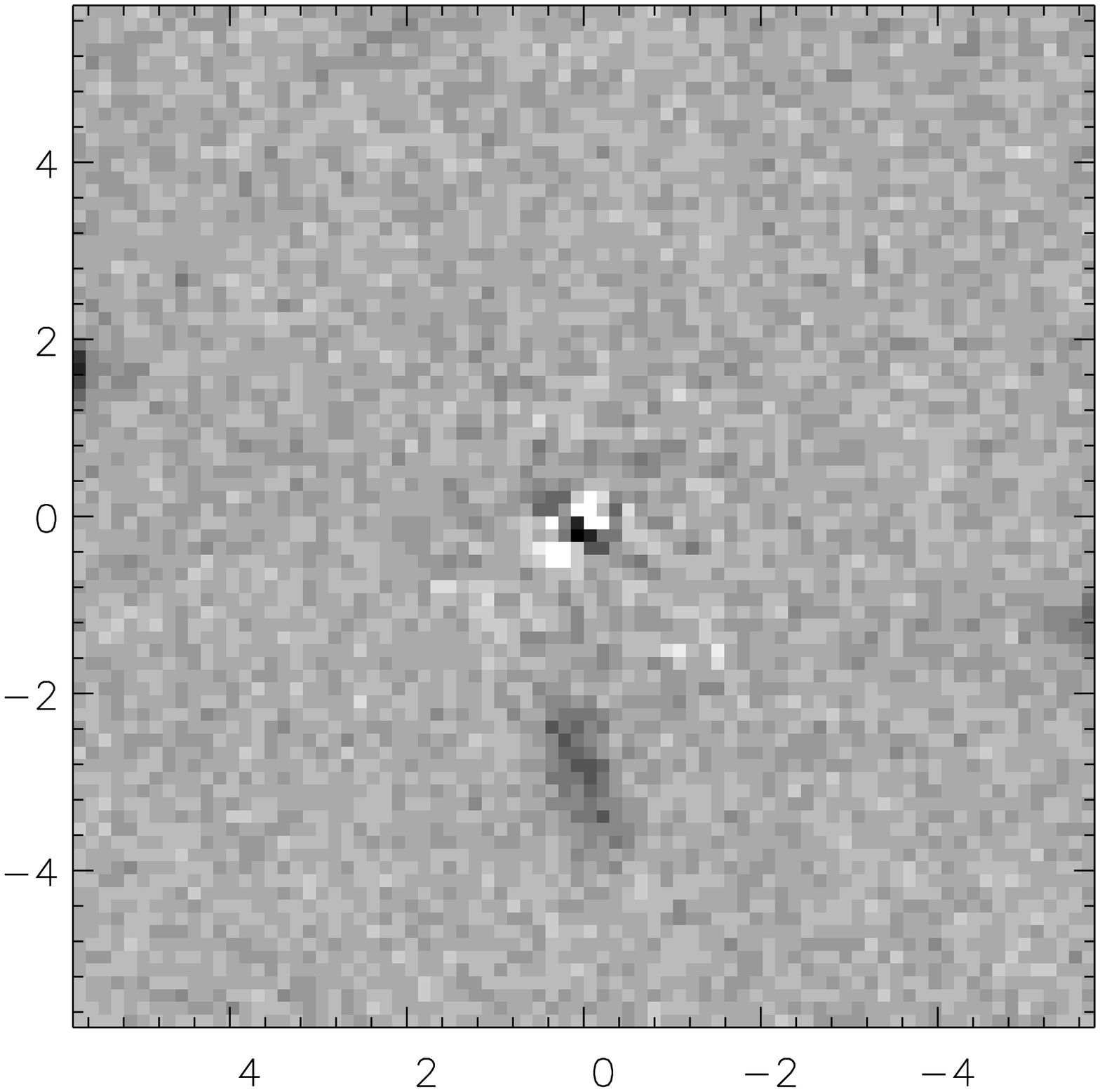,width=0.30\textwidth}\\
  \\

 \end{tabular}

 \caption{The $K_S$-band quasar images, and the results of scaled quasar-PSF subtraction to produce residuals with monotonically declining radial luminosity profiles. The top row shows the image of the quasar J131052.51$-$005533.2 (left), the stellar PSF extracted from within the same image (centre) and the PSF-subtracted monotonically declining host-galaxy residual (right). The second row shows the same information for the quasar SDSS J144617.35$-$010131.1 with its associated on-image stellar PSF. The bottom row shows the effect of performing a control experiment, applying the same analysis to two stellar PSFs drawn from the same science image (left and centre) with the resulting residual from scaled PSF-PSF subtraction, again designed to produce a monotonically-declining residual (right). All panels are $12 \times 12$\,arcsec.}
\end{figure*}

\section{RESULTS}

\begin{table*}
 \begin{center}

 \caption{$K_S$-band photometry of the quasar host galaxies as derived from the scaled PSF-subtracted residuals. Column 1 gives the source name. Column 2 gives the measured 4-arcsec diameter aperture magnitude with associated total error. Columns 3, 4, and 5 list the contributions to this total error provided by uncertainty in sky background, from PSF mismatch, and from uncertainty in the precise level of PSF subtraction required to produce the desired minimum residual image with a monotonically declining radial luminosity profile.}

 \begin{tabular}{lcccc}
 \hline
 Source & $K_S$\,/\,mag & $\delta_K$ back & $\delta_K$ PSF & $\delta_K$ sub\\
 \hline
 SDSS J131052.51$-$005533.2 & 19.42$\pm$0.17 & 0.07 & 0.15 & 0.03\\
 SDSS J144617.35$-$010131.1 & 20.51$\pm$0.29 & 0.12 & 0.26 & 0.07\\
 \hline
 \end{tabular}
 \label{tabphot}

 \end{center}
\end{table*}

\subsection{Host-galaxy luminosities}

In Table 2 we give the measured $K_S$-band aperture magnitudes of the residual host galaxies, along with our estimates of the photometric error contributions from basic uncertainty in the background, PSF mismatch, and uncertainty in the PSF scaling required to produce the monotonically declining residual image.

In an attempt to quantify the extent to which the host galaxy luminosities will have been (inevitably) under-estimated by our adopted process of scaled PSF subtraction, we undertook two-dimensional modelling of the host galaxy residuals shown in Fig. 2 using GALFIT (Peng et al. 2002, 2010). This modelling was also aimed at extracting information on host-galaxy size and morphology, as discussed further below. To minimise the impact on this modelling of the known over-subtraction of the PSF (which of course produces somewhat ``flat-topped'' residual images), the central 9 pixels of each host galaxy residual were excluded from the model fitting. The results of this model fitting are illustrated in Fig. 3, tabulated in Table 3, and discussed further below. 

From the point of view of host galaxy luminosities, the key point is that, as summarised in Table 3, the modelled restoration of additional light in the central regions of the host galaxies does, as expected, make the inferred host galaxies brighter, but only by $\simeq 0.2-0.3$ mag. within our adopted 4-arcsec diameter aperture.

\begin{figure*}
 \begin{tabular}{ccc}

  \epsfig{file=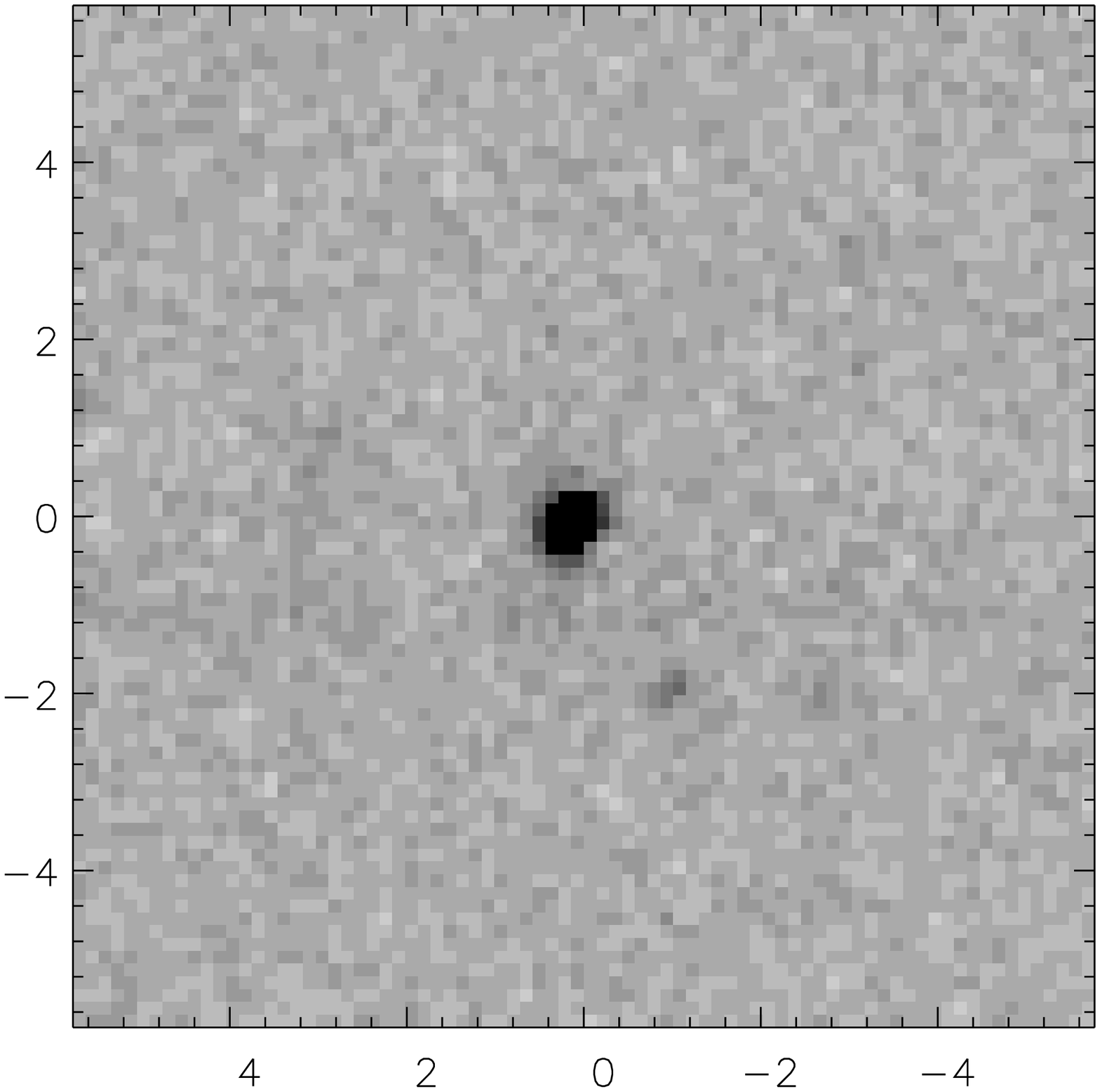,width=0.30\textwidth}&
  \epsfig{file=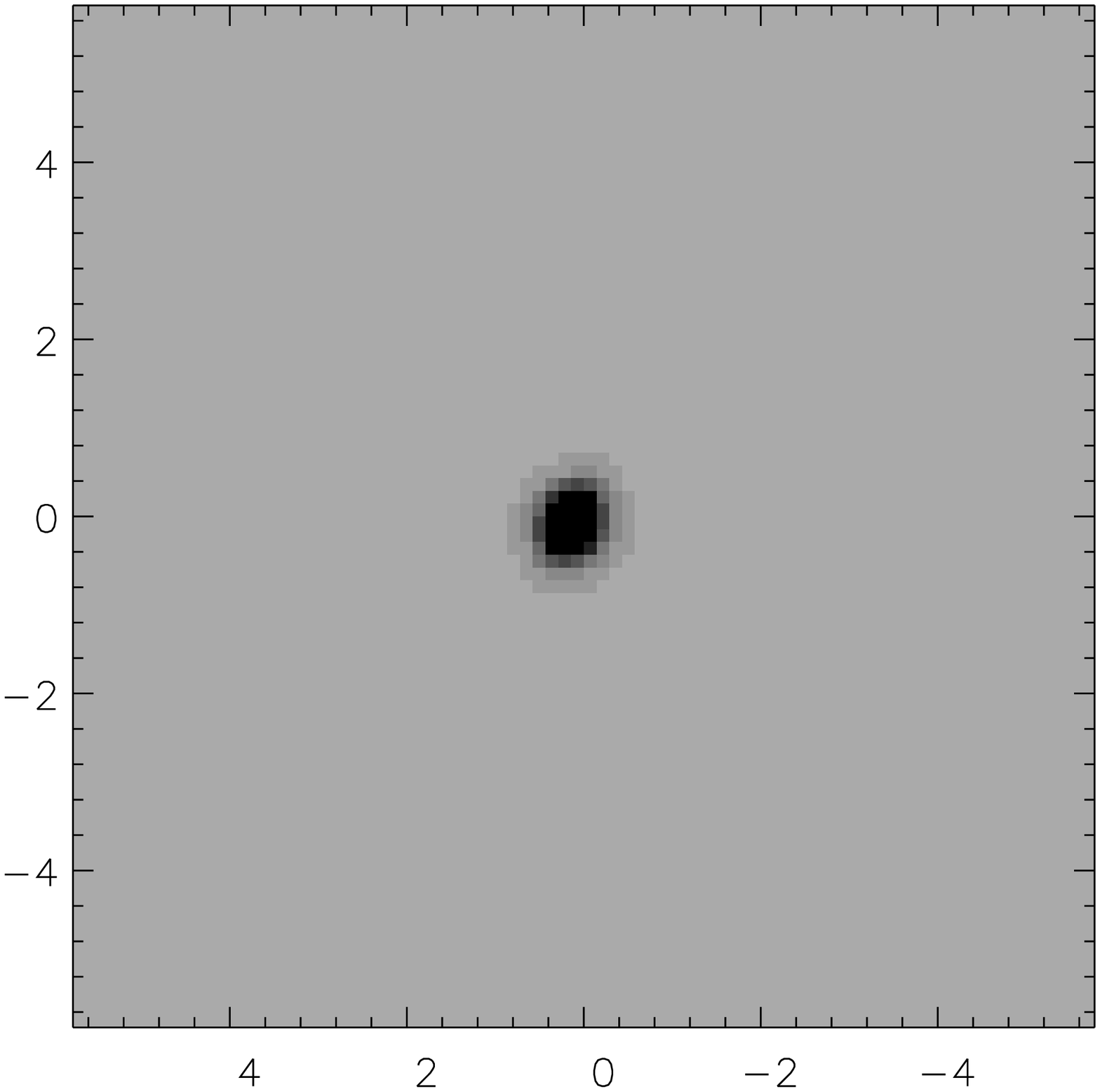,width=0.30\textwidth}&
  \epsfig{file=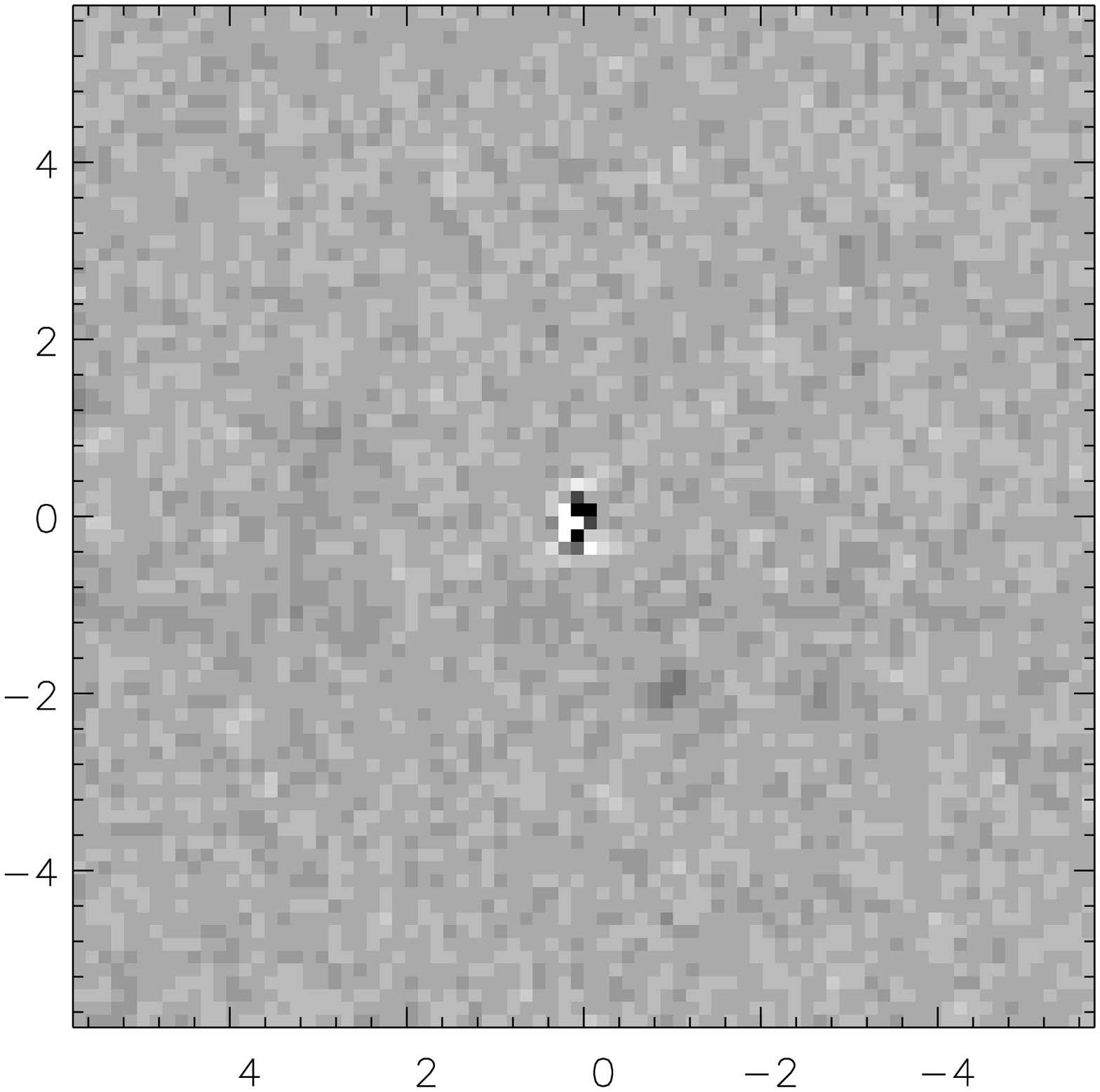,width=0.30\textwidth}\\
  \epsfig{file=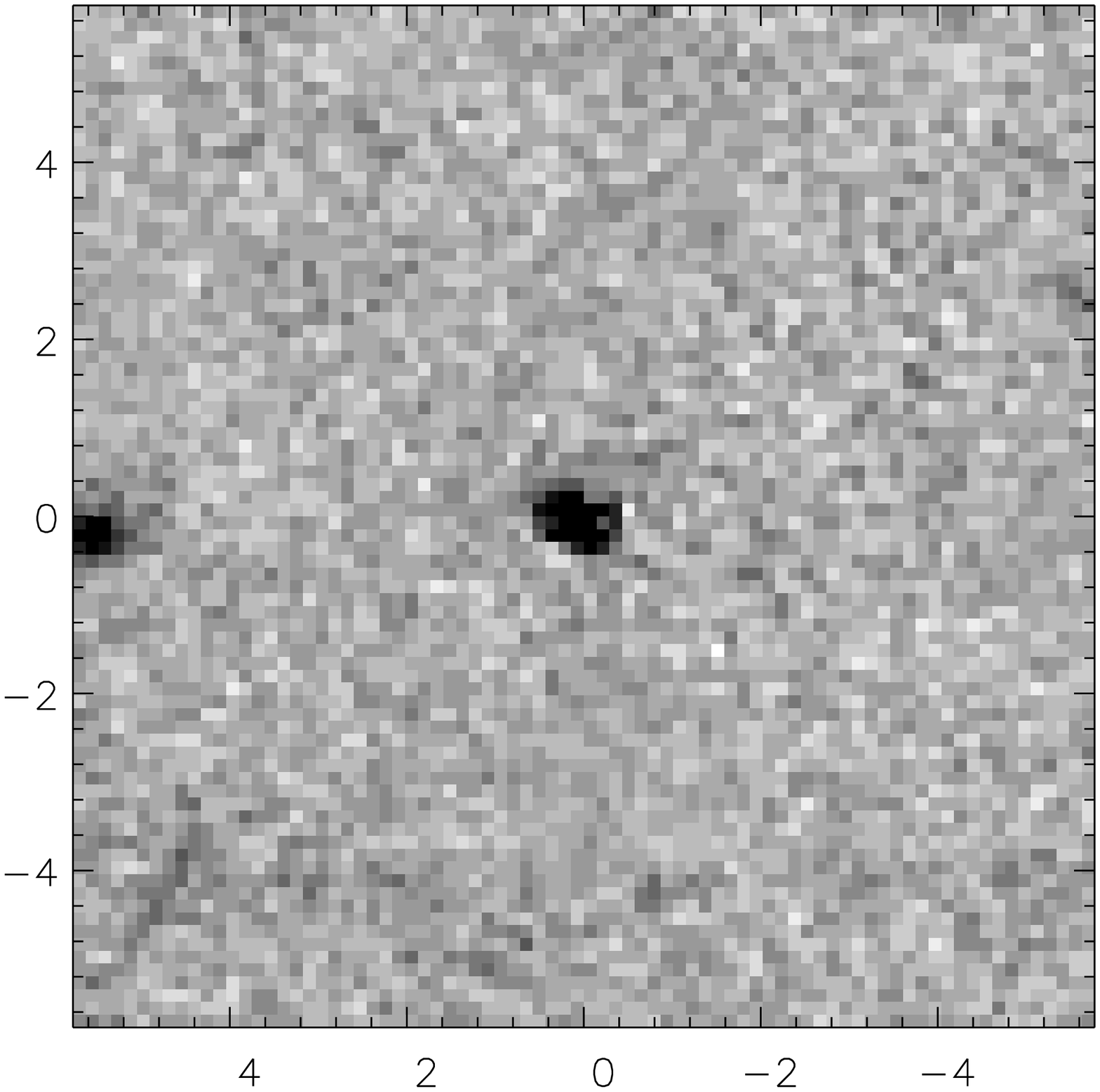,width=0.30\textwidth}&
  \epsfig{file=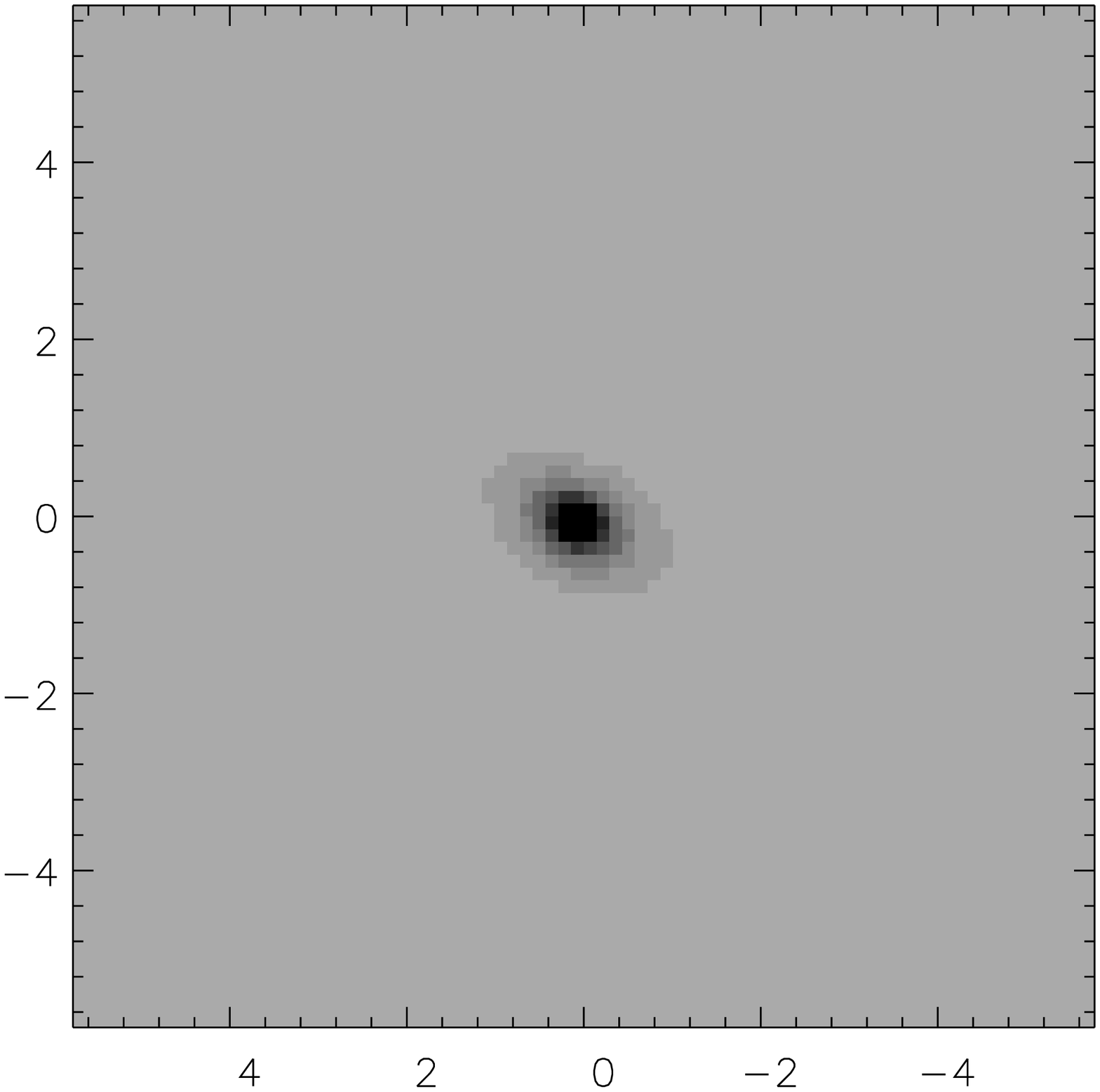,width=0.30\textwidth}&
  \epsfig{file=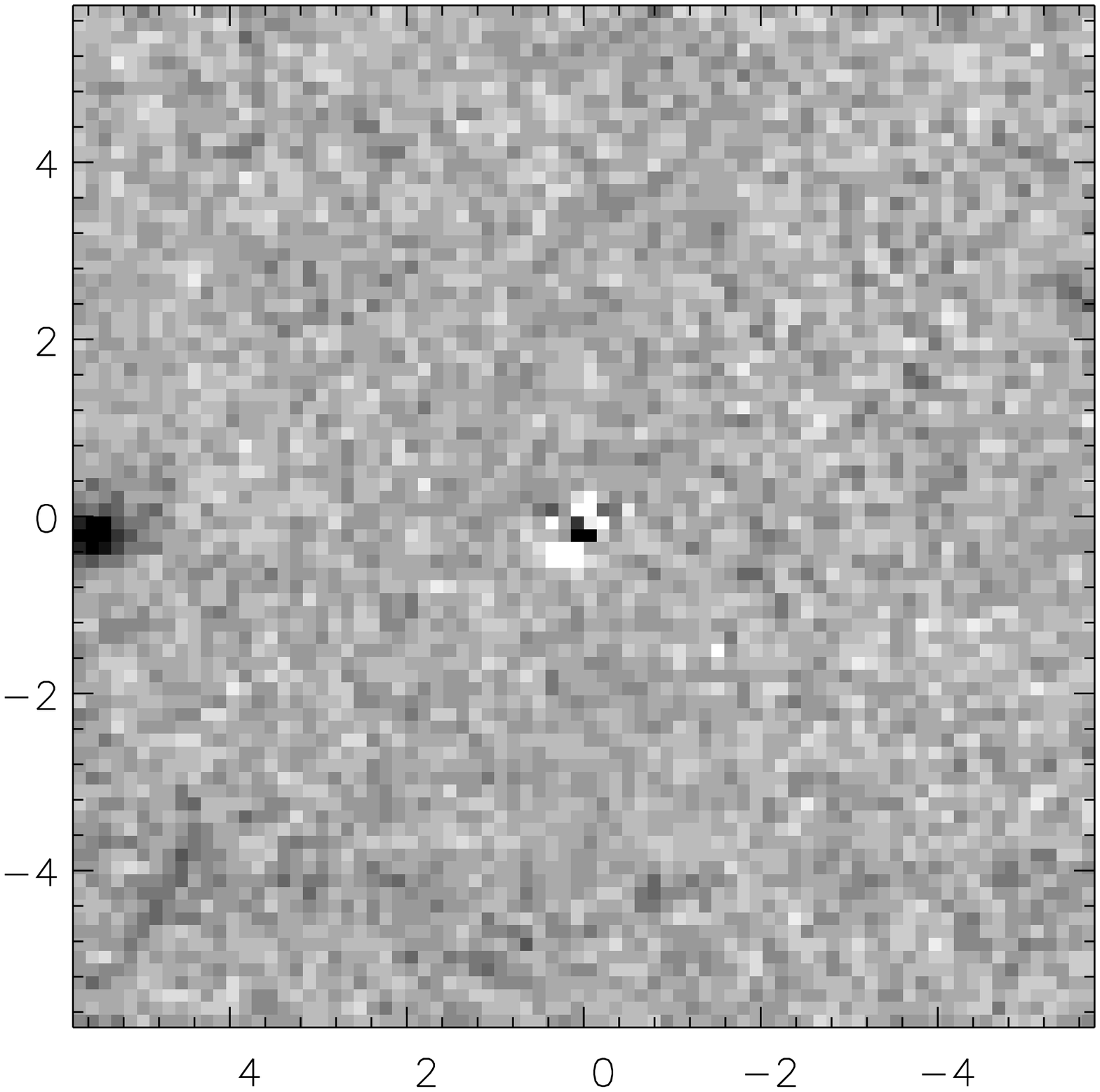,width=0.30\textwidth}\\
  \\

 \end{tabular}

 \caption{Two-dimensional modelling of the host galaxies of SDSS J131052.51$-$005533.2 (upper row) and SDSS J144617.35$-$010131.1 (lower row). The left-hand panel shows the host-galaxy light which remains after subtraction of the scaled PSF to produce the monotonically declining residual. The middle panel shows the best-fitting two-dimensional model after fitting to the residual host galaxy (excluding the central pixels). The right-hand panel shows the final residual after subtraction of the best fitting two-dimensional model from the observed host galaxy. As in Fig.2, all panels are $12 \times 12$\,arcsec.}
\end{figure*}

\subsection{Host-galaxy sizes}

In modelling the quasar host-galaxies with GALFIT we allowed axial ratio, luminosity, scalelength ($r_{1/2}$) and S\'{e}rsic index ($n$) to vary. The resulting best-fitting parameter values are summarised in Table 3, and the resulting model images shown in Fig. 3. The derived best-fitting values of half-light radius and S\'{e}rsic index ($n$) appear perfectly sensible, giving some additional reassurance that we are modelling genuine host-galaxy light. However, we found that while the half-light radius of the host-galaxies appears to be relatively robust, the derived value of S\'{e}rsic index is not. This point is illustrated in Fig. 4 which shows how reduced $\chi^2_{\nu}$ changes as half-light radius is varied, for different alternative fixed values of S\'{e}rsic index $n$. It can be seen that minimum $\chi^2_{\nu}$ is achieved at essentially the same length-scale for each of the three alternative values of $n$ adopted ($n = 2.5, 1, 4$), but that the best quality of fit achieved is essentially independent of $n$. Measurements of raw $\chi^2$ stepping through fixed half-light radii return model parameter uncertainties, with one-sigma errors derived from a $\Delta\chi^2=1$ range in the plot (shown in Table 3). We conclude that we can say nothing about whether the host galaxies of these luminous $z \simeq 4$ quasars are dominated by discs or spheroids, but are reassured that our data are of sufficient quality to constrain their scalelengths and luminosities.

\begin{table*}
 \begin{center}

 \caption{The results of two-dimensional modelling of the residual host-galaxy images. Column 1 gives the source name. Column 2 lists the model semi-major axis scalelength (half-light radius) in kiloparsec. Column 3 lists the value of the S\'{e}rsic index ($n$). Column 4 gives the total model $K_S$-band magnitude. Column 5 gives the model $K_S$-band magnitude within a 4-arcsec diameter aperture (for ease of comparison with the raw residual photometry values given in Table 2). Column 6 gives $\chi^{2}_{\nu}$ for the best-fitting model.} 

 \begin{tabular}{lccccc}
 \hline
 Source & $r_{1/2}$\,/\,kpc & S\'{e}rsic $n$ & $K_S$ model tot & $K_S$ model 4$^{\prime \prime}$ & $\chi^{2}_{\nu}$\\
 \hline
 SDSS J131052.51$-$005533.2 & 0.75$\pm$0.16 & 2.4 & 19.27 & 19.30 & 1.24\\
 SDSS J144617.35$-$010131.1 & 2.84$\pm$0.42 & 2.8 & 20.24 & 20.29 & 1.29\\
 \hline
 \end{tabular}
 \label{tabstack}

 \end{center}
\end{table*}


\begin{figure*}
 \begin{tabular}{cc}

 \epsfig{file=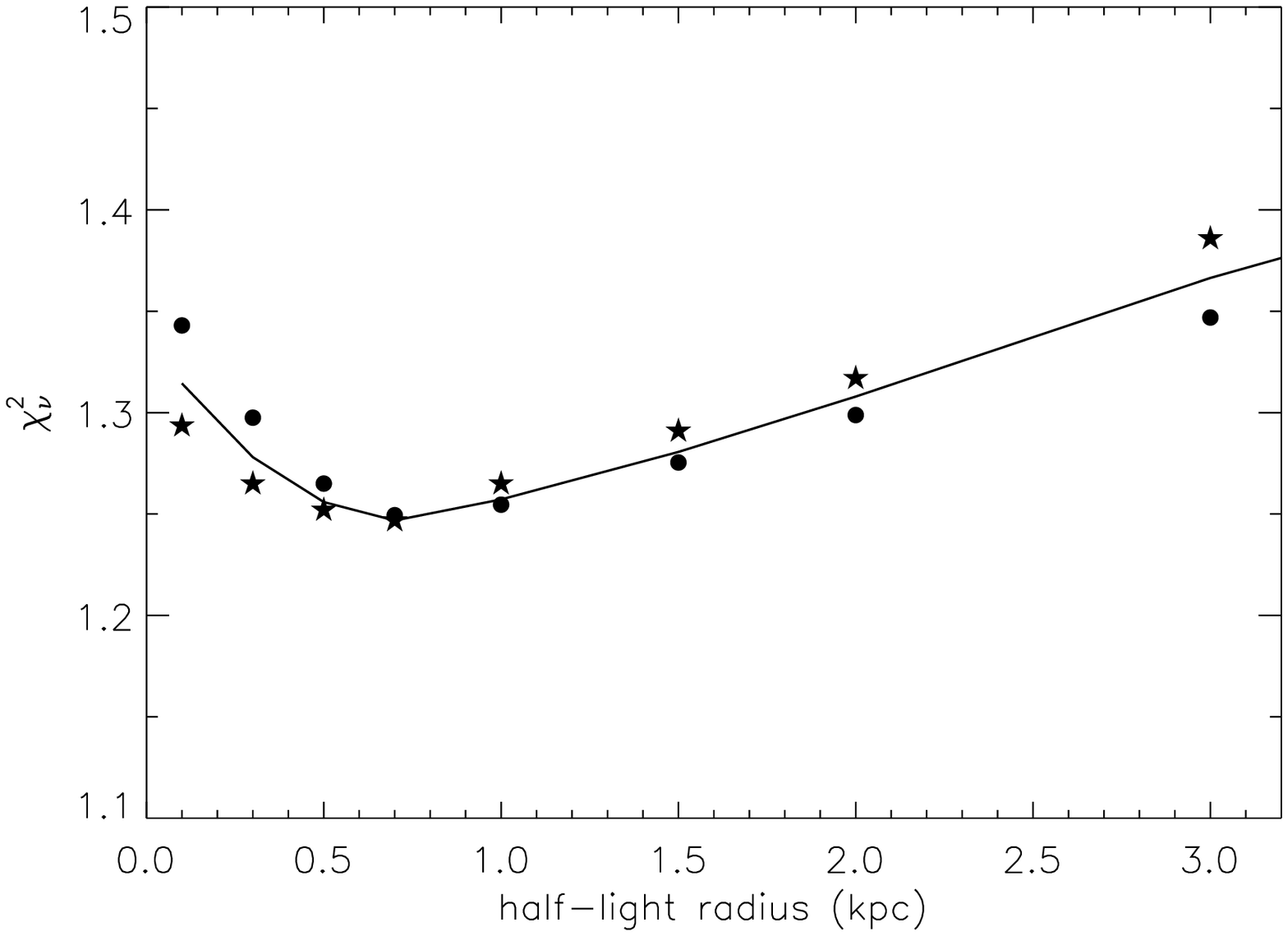,width=0.45\textwidth}&
 \epsfig{file=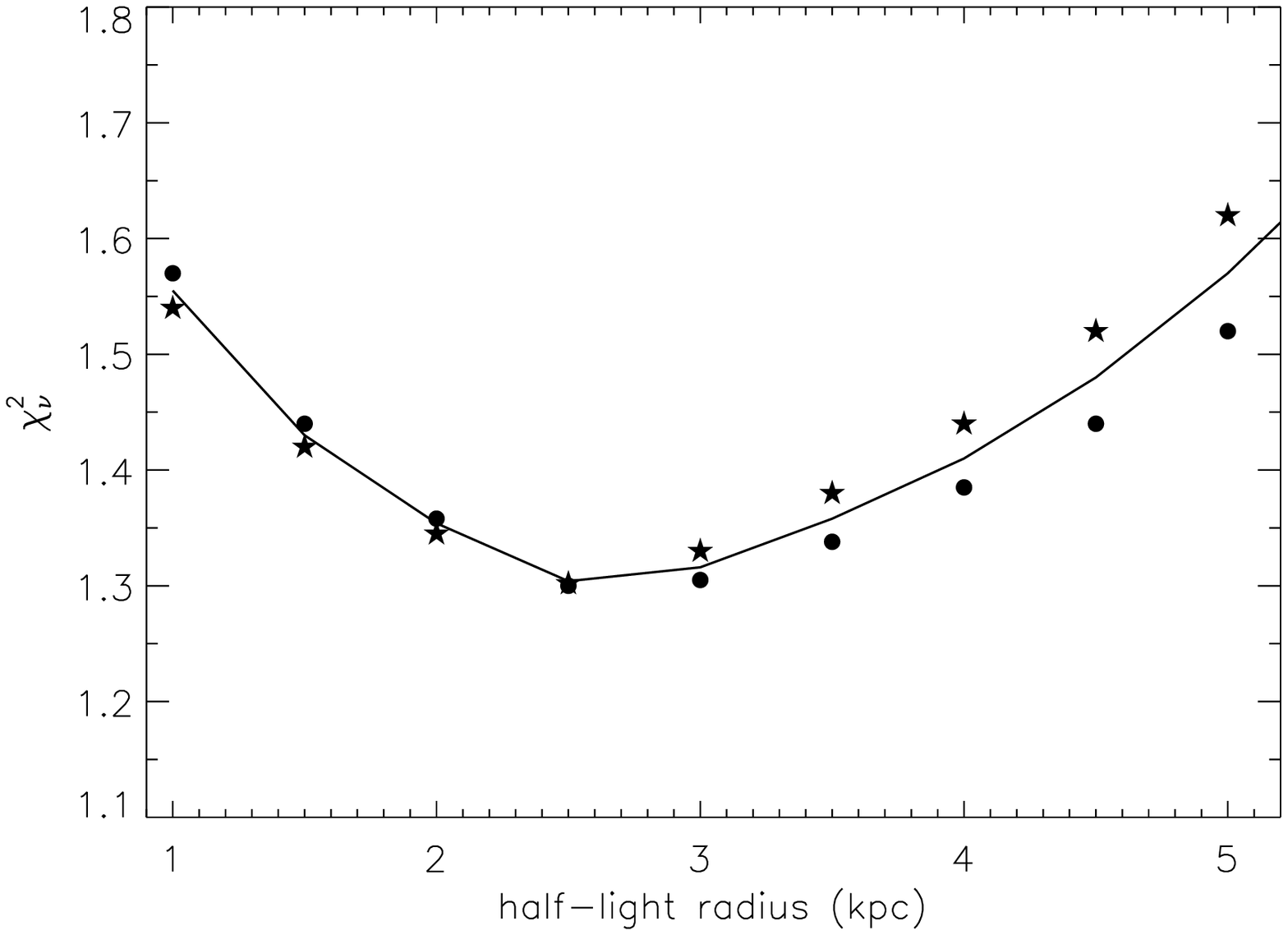,width=0.45\textwidth}\\

 \end{tabular}

 \caption{$\chi^{2}_{\nu}$ plotted against model half-light radius for SDSS J131052.51$-$005533.2 (left) and SDSS J144617.35$-$010131.1 (right) for different adopted morphologies as parameterised by S\'{e}rsic index $n$. Results are shown for fixed values of S\'{e}rsic index $n=1$ (filled stars) and $n=4$ (filled circles), as well as for the average best-fitting value of $n=2.5$ (solid line).}

\end{figure*}

\subsection{Host-galaxy stellar masses}

We have derived estimates of the stellar masses of the two quasar host galaxies using the evolutionary synthesis models of Bruzual \& Charlot (2003), assuming the Initial Mass Function (IMF) of Salpeter (1955). The extent to which an accurate stellar mass can be derived is limited by the fact that, while we do at least know the precise spectroscopic redshift of both quasars, we possess no meaningful information on the colour of either host galaxy. We thus made two simple but extreme estimates of the stellar mass of each host galaxy (effectively bracketing the plausible mass range) by assuming first that the entire galaxy was formed as recently as $z_f = 5$, and then alternatively assuming that the galaxy was formed in a single burst at much higher redshift, $z_f = 10$. While neither assumption is completely realistic, this does at least yield fairly robust upper and lower limits on the possible stellar masses of the two quasar host galaxies. The resulting inferred minimum and maximum stellar masses for each quasar host galaxy are given in Table 4. We note that adoption of the IMF of Chabrier (2003) would yield stellar masses a factor $\simeq 1.8$ lower. 

\begin{table*}
 \begin{center}
 \caption{Estimated galaxy stellar masses and quasar black-hole masses. Column 1 gives the source name. Column 2 gives the host-galaxy stellar mass (in units of $10^9\,{\rm M_{\odot}}$) on the assumption that the entire galaxy was formed in a single starburst at $z_f = 5$, while column 2 gives the largest value of stellar mass inferred from pushing the formation epoch back to $z_f = 10$. Column 4 gives the value of black-hole mass estimated from the SDSS optical spectra as described in Section 4.4.}
 \begin{tabular}{lccc}
 \hline
Source & $M_{gal,min}/{\rm 10^9\,M_{\odot}}$ & $M_{gal,max}/{\rm 10^9\,M_{\odot}}$ & $M_{bh}/{\rm 10^9\,M_{\odot}}$\\
\hline
SDSS J131052.51$-$005533.2 & 300 & 1400 & 9.96\\
SDSS J144617.35$-$010131.1 & 100 & \phantom{1}500 & 9.99\\
 \hline
 \end{tabular}
 \label{masses}
 \end{center}
\end{table*}

\subsection{Black-hole masses}

The masses of the black holes which power these two $z \simeq 4$ quasars were estimated using the so-called virial black-hole mass estimator for broad-line AGN (e.g. Wandel, Peterson \& Malkan 1999; Kaspi et al. 2000). Under the assumption that the broad-line emitting gas is in virial motion within the central black hole's gravitational potential, the central mass can be estimated via $M_{bh}=G^{-1}R_{blr}V^{2}$, where $R_{blr}$ is the radius of the broad-line region (BLR) and $V$ is the orbital velocity of the line-emitting gas. In practice, $R_{blr}$ is estimated via the correlation between optical luminosity and $R_{blr}$ discovered from reverberation mapping of low-redshift AGN (Kaspi et al. 2000), and the gas orbital velocity is taken to be the full width at half-maximum (FWHM) of one or more of the H$\beta$, MgII or CIV emission lines, depending on which lines are observable given the source redshift.

At $z\simeq4$, the CIV line is the only one available within the SDSS optical spectra, and can be used to estimate $M_{bh}$ using the virial black-hole mass estimator from Vestergaard \& Peterson (2006):

$$log \frac{M_{bh}}{\hbox{$\rm\thinspace M_{\odot}$}}=log\left[\left(\frac{\lambda L_{1350}}{10^{44}\,{\rm erg\, s^{-1}}}\right)^{0.53}\left(\frac{FWHM(CIV)}{{\rm 1000\,km\,s}^{-1}}\right)^{2}\right]+6.66$$

The results of this calculation for the two quasars considered here are given in Table 4. It should be noted that there is some concern when using high-ionisation lines as a proxy for BLR rotational velocity, such as CIV, which can be blueshifted with respect to lower-ionisation lines such as H$\beta$ and MgII, and often shows a strong blue excess asymmetry (Baskin \& Laor 2005). This suggests that non-gravitational effects, such as obscuration and radiation pressure, may affect the line profile, and hence distort black-hole masses derived assuming a gravity dominated linewidth. Baskin \& Laor (2005) analyse the H$\beta$ and CIV lines for a sample of 81 PG quasars to conclude that CIV is inferior to H$\beta$ for $M_{bh}$ estimates. However, for CIV lines wider than 4000\,km\,s$^{-1}$ such as the ones in this study, Baskin \& Laor (2005) find that the virial BH mass estimator may actually be biased toward low BH estimates. Any such bias would therefore reinforce the observed evolution discussed in section 5. Conversely, Vestergaard (2009) reanalyse the Baskin \& Laor (2005) data, and the CIV line in a larger sample of AGN, and conclude that CIV is a robust measure BLR rotational velocity. This result is confirmed in Peng et al. (2006b), who find no significant offset between H$\beta$ and CIV linewidths in a sample of six $z\sim1$ quasars. Although possibly inferior to H$\beta$ and MgII, as the only line available at $z>1.5$ it would seem both necessary and acceptable to utilise the CIV line as a proxy for BLR rotational velocity.

\section{DISCUSSION}

In Fig. 5 (adapted from Miley \& De Breuck 2008 following Rocca-Volmerange et al. 2004) we attempt to illustrate how the $K_S$-band brightness of our two $z \simeq 4$ quasar host galaxies compares both with other estimates of quasar host-galaxy brightness at comparable redshift, and with the brightness of other well-studied populations of high-redshift galaxies. Here it can be seen that our quasar hosts lie on the now well-established $K-z$ relation for powerful radio galaxies (e.g. Lilly \& Longair 1984; Eales et al. 1997; van Breugel et al. 1998; De Breuck et al. 2002). As can be seen from Fig. 5, and as inferred from a series of deep near-infrared imaging studies of powerful radio galaxies (e.g. Dunlop \& Peacock 1993; Best et al. 1998; Willott et al. 2003; Targett et al. 2011), the radio-galaxy $K-z$ relation essentially defines the high-mass envelope of the evolving galaxy population. Regardless of their precise stellar masses, it is thus clear that, as perhaps expected, the host galaxies of the most luminous SDSS quasars at $z \simeq 4$ are amongst the most massive known galaxies at this epoch. In addition, while we have argued that our data are of higher quality, and our results more robust than those derived from previous studies of quasar hosts at $z \simeq 4$, our host galaxy luminosities are basically consistent with those derived by McLeod \& Bechtold (2009), Hutchings (2003, 2005) and Peng et al. (2006b). As judged from the galaxy evolution models of Rocca-Volmerage et al. (2004) over-plotted on Fig. 5, it would seem that all of the $z \simeq 4$ quasar host galaxies uncovered to date have stellar masses in the range $1 - 10 \times 10^{11}\,{\rm M_{\odot}}$, which is entirely consistent with our own derived bounds on the possible stellar masses of our two quasar host galaxies as given in Table 4.

\begin{figure*}

\epsfig{file=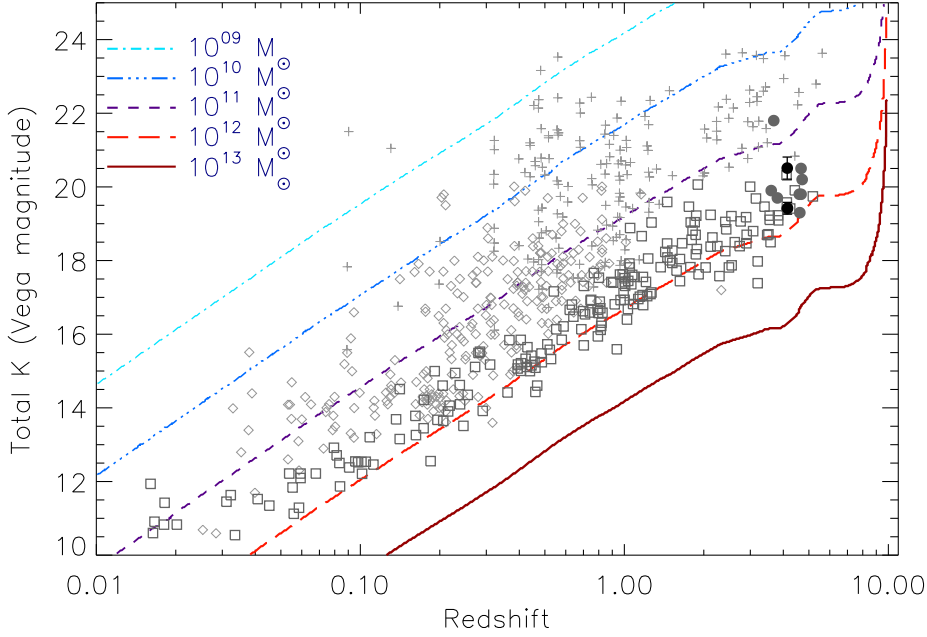,width=0.70\textwidth}

\caption{Composite $K-z$ diagram showing $z\sim4$ quasar hosts, radio galaxies, and optically-selected galaxies, adapted from Miley \& De Breuck (2008) following Rocca-Volmerange et al. (2004). The $z \simeq 4$ SDSS quasar host-galaxy magnitudes derived here are plotted (solid black circles) along with the other $z\sim4$ quasar host-galaxy results (solid grey circles) from the literature (Peng et al. 2006b; Hutchings 2003, 2005; McLeod \& Bechtold 2009). Also plotted are massive radio galaxies (squares) from De Breuck et al. (2002), and optically-selected field galaxies in the Hubble Deep Field North (crosses) and the Hawaii survey (diamonds). These data are over plotted with theoretical $K-z$ tracks for elliptical galaxies formed with a range of assumed initial reservoir gas masses, and developed using P\'{e}gase galaxy evolution models as described in Rocca-Volmerange et al. (2004).}

\end{figure*}

Our study is, however, the first to yield apparently robust values for the half-light radii of quasar host galaxies at such high redshifts. Despite their large stellar masses, the derived half-light radii of the $z \simeq 4$ quasar hosts ($r_{1/2} \simeq 1.8$\,kpc) are clearly much smaller than those of either low-redshift quasar hosts ($\simeq 10$\,kpc; Dunlop et al. 2003) or indeed the vast majority of present-day galaxies of comparable mass (e.g. Hyde \& Bernardi 2009). However, our small derived values for $r_{1/2}$ seem broadly as expected given the rapidly growing observational evidence for the increasing compactness of massive galaxies with increasing redshift (Daddi et al. 2005; Trujillo et al. 2006, 2007; Longhetti et al. 2007; Zirm et al. 2007; Cimatti et al. 2008; Buitrago et al. 2008; van Dokkum et al. 2008; Targett et al. 2011) and the expectations from recent simulations of galaxy growth (e.g. Oser et al. 2012). Our results are also consistent with the only existing indirect size estimates for several $z\sim4$ quasars based on lensed measurements of the molecular gas content (Riechers et al. 2008a,b; Riechers et al. 2009). We conclude that, just as at low redshift, the properties of quasar host galaxies at $z \simeq 4$ seem broadly as expected for ``normal'' galaxies of comparable mass at that particular cosmological epoch.

The derived black-hole masses given in Table 4 are clearly very large, approaching $M_{bh} \simeq 10^{10}\,{\rm M_{\odot}}$ for both quasars, and so it is worth pausing to consider whether such values are plausible/expected. Our conclusion is that these masses, while high, are certainly not unreasonable given the relative rarity of these ultra-bright quasars, our understanding of quasar demographics, and the results of the latest high-redshift simulations. Specifically, our target objects were selected from a parent sample which contains only 1 such luminous quasar in the redshift range $4 < z < 5$ per 50 deg$^2$ on the sky, corresponding to a comoving source density of only $\simeq 2\,{\rm Gpc^{-3}}$. These are thus extreme objects, arguably the natural descendents of the most extreme quasars found at $ z \simeq 6.5 - 7$, which already appear to have black-hole masses $M_{bh} \simeq 2 \times 10^9\,{\rm M_{\odot}}$ (Willott, McLure \& Jarvis 2003; Mortlock et al. 2011). Only 60 million years of further Eddington-limited growth would be required to boost these black-hole masses by a further factor of 4, to values consistent with those derived here 800 million years later. This viewpoint receives some further theoretical support from recent simulations of black-hole growth in the young Universe, such as the ``Massive Black'' simulation of Di Matteo et al. (2011). The luminous quasars targeted here are sufficiently rare that, on average, only one would be expected in the $\simeq 0.5\,{\rm Gpc^3}$ comoving volume followed in this simulation, and Di Matteo et al. report that the most massive black hole in their simulation has already achieved a mass of $M_{bh} \simeq 5 \times 10^9\,{\rm M_{\odot}}$ by $z \simeq 5$. By this point feedback and shock heating of infalling gas are inferred to limit further growth, but clearly the concept that such rare black holes may achieve masses approaching $M_{bh} \simeq 10^{10}\,{\rm M_{\odot}}$ by $z \simeq 4$ is consistent with current observational and theoretical constraints.

\begin{figure*}

\epsfig{file=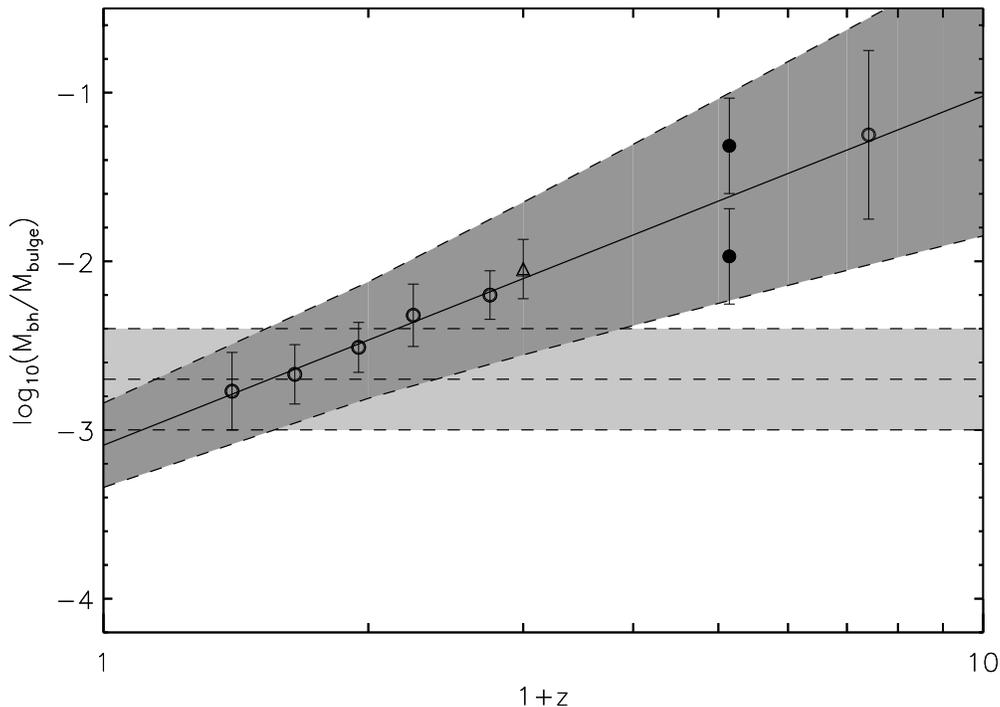,width=0.82\textwidth}

\caption{Average $M_{bh}:M_{gal}$ of the $z \simeq 4$ SDSS quasars (filled circles) for both the maximum (lower point) and minimum (upper point) galaxy masses, derived for the host galaxies where formation redshift was constrained to be $z_f=10$ or $z_f=5$ respectively. The solid line shows the best fit to the observed evolution in 3CRR galaxies (open circles) from McLure et al. (2006), while the dark-grey shaded area shows the 1$\sigma$ uncertainty of this fit. This trend is consistent with results from Peng et al. (2006b), who find a similar evolution out to $z\sim2$ in a sample of 11 quasars (open triangle). The light-grey shaded area illustrates the $\pm0.3$-dex uncertainty on the local $M_{bh}/M_{bulge}$ ratio, centred on $M_{bh}/M_{bulge} = 0.002$. The data point at $z = 6.41$ from Willott, McLure, \& Jarvis (2003) is based not on any estimate of the stellar mass of the host galaxy, but on a dynamical mass inferred from Very Large Array measurements of molecular gas in the quasar host galaxy by Walter et al. (2004).}

\end{figure*}

Finally, we can combine our estimates of black-hole mass and galaxy stellar mass to infer the black hole-to-galaxy mass ratio for our $z \simeq 4$ quasars. The averaged value of $M_{bh}:M_{gal}$ is shown in Fig. 6, where we plot two $z\simeq4$ data points to indicate the different values which arise from adoption of either the maximum or minimum possible galaxy stellar masses (as derived assuming a formation redshift of $z_f=10$ or $z_f=5$ respectively). The light-grey shaded area illustrates the $\pm0.3$-dex uncertainty on the local $M_{bh}/M_{bulge}$ ratio, centred on $M_{bh}/M_{bulge} = 0.002$. The solid line shows the best fit to the observed evolution in 3CRR galaxies from McLure et al. (2006), while the dark-grey shaded area shows the 1$\sigma$ uncertainty of this fit. This evolving $M_{bh}/M_{bulge}$ ratio result is consistent with results from Peng et al. (2006a), who find a similar evolution out to $z\sim2$ in a sample of 11 quasars. Also shown is the $z=6.41$ quasar observed by Willott, McLure, \& Jarvis (2003) who used $H$-band and $K$-band spectra covering the MgII broad emission line to derive a black-hole mass of $3\times10^{9}\,{\rm M_{\odot}}$. However, note that in this case the host-galaxy mass refers not to a stellar mass inferred from the sort of host-galaxy imaging attempted here, but is a dynamical mass inferred from Very Large Array measurements of molecular gas in the quasar host galaxy by Walter et al. (2004). Even given the uncertainties, our results imply a black-hole:host-galaxy mass ratio ($M_{bh}:M_{gal} \simeq 0.01 - 0.05$) which appears to be a factor of $\simeq 10$ higher than typically seen in the low-redshift Universe. Fig. 6 demonstrates that this is consistent with the growing body of evidence for a systematic growth in this mass ratio with increasing redshift, at least for objects selected as powerful active galactic nuclei (Decarli et al. 2010, and references therein). Indeed, an order of magnitude increase in $M_{bh}:M_{gal}$ is exactly as anticipated given the redshift dependence of $M_{bh}:M_{gal} \propto (1 + z)^{1.4}$ inferred from recent studies of X-ray selected AGN (Bennert et al. 2011), and recent numerical simulations of galaxy and black hole growth (e.g. Shankar et al. 2011, Power et al. 2011).

The observed evolution in the black-hole:host galaxy mass ratio shown in Fig. 6, although consistent with numerous results in the literature, could be subject to various selection effects. The necessity of using black-hole masses derived from the velocity width of broad permitted lines limits candidate objects to luminous rapidly-accreting unobscured Type-1 active galactic nuclei, which implies that results will necessarily be biased from the true distribution of a volume-limited sample (e.g. Lauer et al. 2007). For example, the low $M_{bh}:M_{gal}$ estimated for submillimetre selected galaxies at $z\sim2.2$ (Borys et al. 2005) could imply a large scatter in the relationship at high redshift not sampled by quasars. However, the agreement between our results and other studies using both quasars and radio galaxies indicates that, at least for the most massive galaxies, the growth of the central supermassive black hole may be completed before that of the host spheroid.

\section{CONCLUSION}

We have obtained and analysed the deepest, high-quality $K_S$-band images ever obtained of luminous quasars at $z \simeq 4$, in an attempt to determine the basic properties of their host galaxies less than 1 Gyr after the first recorded appearance of black holes with $M_{bh}~>~10^9\,{\rm M_{\odot}}$. To maximise the robustness of our results, we carefully selected two SDSS quasars at $z \simeq 4$. With absolute magnitudes $M_i < -28$, these quasars are representative of the most luminous quasars known at this epoch but they also, crucially, lie within 40 arcsec of comparably-bright foreground stars (required for accurate point-spread-function definition), and have redshifts which ensure line-free $K_S$-band imaging. 

Our new data were obtained in excellent seeing conditions ($<0.4$-arcsec) with ISAAC on the ESO VLT, with integration times of $\simeq 5.5$ hours per source. Via carefully-controlled separation of host-galaxy and nuclear light, followed by two-dimensional galaxy model-fitting, we have estimated the luminosities and stellar masses of the host galaxies. In addition, although it did not prove possible to determine basic morphological type (e.g. S\'{e}rsic index $n$), we have extracted what appear to be robust estimates of the host-galaxy half-light radii.

We find that the $K_S$-band magnitudes of the quasar host galaxies are consistent with those of luminous radio galaxies at comparable redshifts. This suggests that these quasar hosts are also among the most massive galaxies in existence at this epoch, and indeed inferred stellar masses of these host galaxies lie in the range $2 - 10 \times 10^{11}\,{\rm M_{\odot}}$, depending on assumed formation redshift. We also find that these massive $z \simeq 4$ quasar host galaxies are a factor $\sim 5$ smaller ($\langle r_{1/2} \rangle = 1.8\,{\rm kpc}$) than the host galaxies of luminous low-redshift quasars, or indeed quiescent galaxies of comparable stellar mass. However, we argue that this is as expected given the growing evidence for the increasing compactness of massive galaxies with increasing redshift, and conclude that, just as at low redshift, the properties of quasar host galaxies at $z \simeq 4$ are broadly as expected for ``normal'' galaxies of comparable mass at the relevant cosmological epoch.

Finally we have used the CIV emission line in the SDSS optical spectra of the quasars to estimate the masses of their central supermassive black holes. The results imply extreme black-hole masses approaching $M_{bh} \simeq 10^{10}\,{\rm M_{\odot}}$, but we argue that this is not unrealistic given the extreme nature and rarity of our target $z \simeq 4$ SDSS quasars (i.e. $1 - 2$ per comoving Gpc$^3$), and current observational and theoretical understanding of black-hole growth in the young Universe. Combining the black-hole and host-galaxy mass measurements we infer a black-hole:host-galaxy mass ratio $M_{bh}:M_{gal} \simeq 0.01 - 0.05$. This is an order of magnitude higher than typically seen in the low-redshift Universe, and is consistent with existing evidence for a systematic growth in this mass ratio with increasing redshift, at least for objects selected as powerful active galactic nuclei.

\section*{ACKNOWLEDGEMENTS}
TAT acknowledges support from the European Research Council. JSD acknowledges the support of the Royal Society via a Wolfson Research Merit award, and also the support of the European Research Council via the award of an Advanced Grant. RJM acknowledges the support of the Royal Society via a University Research Fellowship. This work is based primarily on observations made with the Very Large Telescope at the ESO Paranal Observatory under Programme ID 077.B-0358(A).

{}


\begin{thebibliography}{}
\bibitem{1} Baskin A., Laor A., 2005, MNRAS, 356, 1029
\bibitem{2} Bennert V.N., Auger M.W., Treu T., Woo J.-H., Malkan M.A., 2011, ApJ, 742, 107
\bibitem{3} Best P.N., Longair M.S., R{\"o}ttgering H.J.A., 1998, MNRAS, 295, 549
\bibitem{4} Bettoni D., Falomo R., Fasano G., Govoni F., 2003, A\&A, 399, 869
\bibitem{5} Borys C., Smail I., Chapman S.C., Blain A.W., Alexander D.M., Ivison R.J., 2005, ApJ, 635, 853
\bibitem{6} Bruzual G., Charlot S., 2003, MNRAS, 344, 1000
\bibitem{7} Buitrago F., Trujillo I., Conselice C.J., Bouwens R.J., Dickinson M., Yan H., 2008, ApJ, 687, L61
\bibitem{8} Chabrier G., 2003, PASP, 115, 763
\bibitem{9} Cimatti A., et al., 2008, A\&A, 482, 21
\bibitem{10} Daddi E., et al., 2005, ApJ, 626, 680
\bibitem{11} De Breuck C., van Breugel W., Stanford S.A., R{\"o}ttgering H., Miley G., Stern D., 2002, AJ, 123, 637
\bibitem{12} Di Matteo T., Khandi N., DeGraf C., Feng Y., Croft R.A.C., Lopez J., Springel V., 2011, ApJ, submitted, arXiv:1107.1253
\bibitem{13} Decarli R., Falomo R., Treves A., Labita M., Kotilainen J.K., Scarpa R., 2010, MNRAS, 402, 2453
\bibitem{14} Disney M.J., et al., 1995, Nature, 376, 150
\bibitem{15} Dunlop J.S., Peacock J.A., 1993, MNRAS, 263, 936
\bibitem{16} Dunlop J.S., McLure R.J., Kukula M.J., Baum S.A., O'Dea C.P., Hughes D.H., 2003, MNRAS, 340, 1095
\bibitem{17} Eales S., Rawlings S., Law-Green D., Cotter G., Lacy M., 1997, MNRAS, 291, 593
\bibitem{18} Fan X., et al., 2001, AJ, 122, 2833
\bibitem{19} Fan X., et al., 2003, AJ, 125, 1649
\bibitem{20} Floyd D.J.E., Kukula M.J., Dunlop J.S., McLure R.J., Miller L., Percival W.J., Baum S.A., O'Dea C.P., 2004, MNRAS, 355, 196
\bibitem{21} Gebhardt K., et al., 2000, ApJ, 539, L13
\bibitem{22} Ho L.C., 2002, ApJ, 564, 120
\bibitem{23} Hutchings J.B., 2003, AJ, 125, 1053
\bibitem{24} Hutchings J.B., 2005, PASP, 117, 1250
\bibitem{25} Hyde J.B., Bernardi M., 2009, MNRAS, 394, 1978
\bibitem{26} Jahnke K., Maccio A.V., 2011, ApJ, 734, 92
\bibitem{27} Kaspi S., et al., 2000, ApJ, 533, 631
\bibitem{28} Kormendy J., Richstone D., 1995, ARAA\&A, 33, 581
\bibitem{29} Kukula M.J., Dunlop J.S., McLure R.J., Miller L., Percival W.J., Baum S.A., O'Dea C.P., 2001, MNRAS, 326, 1533
\bibitem{30} Lauer T.R., Tremaine S., Richstone D., Faber S.M., 2007, ApJ, 670, 249
\bibitem{31} Lilly S.J., Longair M.S., 1984, MNRAS, 211, 833
\bibitem{32} Longhetti M., et al., 2007, MNRAS, 374, 614
\bibitem{33} Magorrian J., et al., 1998, AJ, 115, 2285
\bibitem{34} Marconi A., Hunt L.K., 2003, ApJ, 589, L21
\bibitem{35} McLeod K.K., Bechtold J., 2009, ApJ, 704, 415
\bibitem{37} McLure R.J., Dunlop J.S., 2001, MNRAS, 327, 199
\bibitem{38} McLure R.J., Dunlop J.S., 2002, MNRAS, 331, 795
\bibitem{39} McLure R.J., Jarvis M.J., 2002, MNRAS, 337, 109
\bibitem{40} McLure R.J., Dunlop J.S., 2004, MNRAS, 352, 1390
\bibitem{41} McLure R.J., Kukula M.J., Dunlop J.S., Baum S.A., O'Dea C.P., Hughes D.H., 1999, MNRAS, 308, 377
\bibitem{42} McLure R.J., Jarvis M.J., Targett T.A., Dunlop J.S., Best P.N., 2006, MNRAS, 368, 1395
\bibitem{43} Merloni A., et al., 2010, ApJ, 708, 137
\bibitem{44} Merritt D., Ferrarese L., 2001, MNRAS, 320, L30
\bibitem{45} Miley G., De Breuck C., 2008, A\&ARv, 15, 67
\bibitem{46} Mortlock D.J., et al., 2011, Nature, 474, 616
\bibitem{47} Oser L., Naab T., Ostriker J.P., Johansson P.H., 2012, ApJ, 744, 63
\bibitem{48} Peng C.Y., Ho L.C., Impey C.D., Rix H.-W., 2002, AJ, 124, 266
\bibitem{49} Peng C.Y., Impey C.D., Ho L.C., Bartob E.J., Rix H.-W., 2006a, ApJ, 640, 114
\bibitem{50} Peng C.Y., Impey C.D., Rix H.-W., Kochanek C.S., Keeton C.R., Falco E.E., Lehar J., McLeod B.A., 2006b, ApJ, 649, 616
\bibitem{51} Peng C.Y., Ho L.C., Impey C.D., Rix H.-W., 2010, AJ, 139, 2097
\bibitem{52} Power C., Zubovas K., Nayakshin S., King A.R., 2011, MNRAS, 413, 1.
\bibitem{53} Rocca-Volmerange B., Le Borgne D., De Breuck C., Fioc M., Moy E., 2004, A\&A, 415, 931
\bibitem{54} Ridgway S.E., Heckman T.M., Calzetti D., Lehnert M., 2001, ApJ, 550, 122
\bibitem{55} Riechers D.A., Walter F., Brewer B.J., Carilli C.L., Lewis G.F., 2008a, ApJ, 686, 851
\bibitem{56} Riechers D.A., Walter F., Carilli C.L., Bertoldi F., Momjian E., 2008b, ApJL, 686, L9
\bibitem{57} Riechers D.A., Walter F., Carilli C.L., Lewis G.F., 2009, ApJ, 690, 463
\bibitem{58} Salpeter E.E., 1955, ApJ, 121 161
\bibitem{59} Schneider D.P., et al., 2010, AJ, 139, 2360
\bibitem{59} Shankar F., Marulli F., Bernardi M., Mei S.,  Meert A.,  Vikram V., 2011, MNRAS, in press (arXiv:1105.6043)
\bibitem{60} Silk J., Rees M.J., 1998, A\&A, 331, L1
\bibitem{61} Szomoru D., et al., 2010, ApJL, 714, L244
\bibitem{62} Targett T.A., Dunlop J.S., McLure R.J., Best P.N., Cirasuolo M., Almaini O., 2011, MNRAS, 412, 295
\bibitem{63} Tremaine S., et al., 2002, ApJ, 574, 740
\bibitem{64} Trujillo I., et al., 2006, MNRAS, 373, L36
\bibitem{65} Trujillo I., Conselice C.J., Bundy K., Cooper M.C., Eisenhardt P., Ellis R.S., 2007, MNRAS, 382, 109
\bibitem{66} Walter F., Carilli C., Bertoldi F., Menten K., Cox P., Lo K.Y., Fan X., Strauss M.A., 2004, ApJ, 615, L17
\bibitem{67} Wandel A., Peterson B.M., Malkan M.A., 1999, ApJ, 526, 579
\bibitem{68} Willott C.J., McLure R.J., Jarvis M.J., 2003, ApJ, 587, 15
\bibitem{69} Willott C.J., et al., 2007, AJ, 134, 2435
\bibitem{70} Willott C.J., et al., 2010, AJ, 140, 546
\bibitem{71} van Breugel W.J.M., Stanford S.A., Spinrad H., Stern D., Graham J.R., 1998, ApJ, 502, 614
\bibitem{72} van Dokkum P.G., et al., 2008, ApJ, 677, 5
\bibitem{73} Vestergaard M., Peterson B.M., 2006, ApJ, ApJ, 641, 689
\bibitem{74} Vestergaard M., Fan X., Tremonti C.A., Osmer P.S., Richards G.T., 2008, ApJ, 674, L1
\bibitem{75} Vestergaard M., 2009, CUP, in press (arXiv:0904.2615)
\bibitem{76} Zirm A.W., et al., 2007, ApJ, 656, 66
\end{thebibliography}
\end{document}